# Emergence of quantum Griffiths singularity in disordered TiN thin films

Sachin Yadav,* [1,2] M.P. Saravanan, [3] and Sangeeta Sahoo* [1, 2]

[1]CSIR-National Physical Laboratory, Dr. K. S. Krishnan Marg, New Delhi-110012, India

[2]Academy of Scientific and Innovative Research (AcSIR), Ghaziabad- 201002, India

[3]Low Temperature Laboratory, UGC-DAE Consortium for Scientific Research, University Campus, Khandwa Road, Indore 452001, India

*Correspondences should be addressed to S. Y. (Email: sachin5459yadav@gmail.com) and

S. S. (Email: sahoos@nplindia.org)

**Abstract**

The association of quantum Griffiths singularity (QGS) to the magnetic-field-induced superconductor-metal transition predicts the unconventional diverging behaviour of dynamical critical exponent in low disorder crystalline two-dimensional superconductors. But whether this state exists in the superconducting systems exhibiting superconductor-insulator transition remains elusive. Here, we report the emergence of quantum Griffiths singularity in ultrathin disordered TiN thin films with more than two orders of magnitude variation in their normal state resistance. For both superconductor-metal transition and superconductor-insulator transition types, a diverging critical exponent is observed while approaching the quantum phase transition. Further, the magnetoresistance isotherms obey a direct activated scaling governed by an infinite-randomness fixed critical point. Finally, this work establishes the robustness of the QGS phenomenon towards a wide range of temperature and also towards a wide range of disorder strength as correlated with the normal state resistance.

# Introduction

Quantum phase transitions (QPTs) corresponding to superconductor-insulator transition (SIT) and superconductor-metal transition (SMT) in two-dimensional (2D) superconductors have been a subject of great interest in last few decades [1,2][3]. QPTs can be initiated by tuning non-thermal external control parameters such as disorder (through film thickness variations) [4], magnetic field [5], carrier concentrations [6], pressure among others[7]. There exists a quantum critical point (QCP) associated with a continuous QPT which separates the superconducting ground state to the normal insulating/metallic state of the system. In the immediate vicinity of a continuous QPT, a $d$-dimensional quantum system can be represented by a ($d$+1) dimensional classical system with imaginary time as the extra dimension. Upon approaching the QCP, both correlation length and correlation time diverge. The divergence can be expressed using characteristic length scales in both spatial ($\xi$) and temporal directions ($\xi_\tau$) as, $\xi \sim |\delta|^{-\nu}, \xi_\tau \sim \xi^z \sim |\delta|^{-z\nu}$, where $\delta$ measures the distance from the critical point. The exponents $\nu$ and $z$ correspond to the correlation length critical exponent and the dynamical critical exponent, respectively. Thus, a continuous QPT leads to a power-law singularities in the length scales at the QCP indicating a power-law singularities in observables too [8] and the corresponding scaling of physical quantities represent universal behavior in the critical exponents $z\nu$ [7].

However, in contrast to the QPT associated with a single QCP in the clean limit, the emergence of multiple critical points (MCPs) in magnetic-field driven QPT has been observed in various systems [9,10] where the disorder plays a major role in controlling the phase transition. The presence of spatial quenched disorders in the form of impurities, defects, dislocations, grain boundaries and others are inevitable in any real sample and assessing their influence on QPT in $d$-dimensional system is generally done by the Harris criterion, $d\nu>2$ with $\nu$ being the correlation length exponent [11]. On large length scale, when the effective disorder strength and its randomness average out under course graining, the Harris criterion is fulfilled and the system undergoes a pure fixed critical point transition with clean limit critical exponents. In contrast, the quenched disorder becomes relevant for the systems that violate the Harris criterion [12]. When the disorder strength reaches a finite value upon course graining, the randomness becomes competitive at all scale and the transition is controlled by a random fixed critical point with critical exponents ($z'\nu'$) differing from those of the clean limit ($z\nu$). Here, the modified correlation length exponent ($\nu'$) fulfils the Harris criterion and the transition resembles the conventional clean critical behavior and the dynamical critical exponent ($z'$) saturates at a disorder dependent finite value [12,13]. In this case, the physical properties of the system are dominated by the rare regions that are spatially localized by the influence of the quenched disorder. Further, an extreme case appears when the disorder strength diverges at the transition under course graining, and the Harris criterion is violated. The transition here is governed by a fixed critical point which is of infinitely strong randomness with diverging dynamical critical exponent. Here, the characteristic length scales in the temporal ($\xi_\tau$) and spatial direction ($\xi$), in contrast to the clean limit power-law scaling ($\xi_\tau \sim \xi^z$), follow an activated scaling as $\xi_\tau \sim e^{\xi^\psi}$ with $\psi$ being the tunneling correlation exponent. This extreme case of quenched disorder mediated criticality of infinite randomness with diverging dynamical critical exponent at the transition is known as quantum Griffiths singularity (QGS) [14]. Therefore, the systematic ways to establish the existence of QGS include the appearance of a continuum of MCPs in field dependent resistance $R_s(B)$ at low temperature, divergence of dynamical exponent at zero-temperature limit while approaching the QPT and the applicability of an activated scaling to describe the physical quantities near the transition.

In 2D superconductors, the first evidence of QGS came from molecular-beam-epitaxy (MBE) grown Ga trilayer [15] which was then followed by other low dimensional superconducting systems [16-26]. The observation of MCPs near the field induced QPTs in these systems led to the emergence of QGS at very low temperature above the mean field upper critical field $B_{c_2}^{MF}$ [15,25,27]. The QGS can be originated by quenched disorder mediated formation of rare ordered superconducting puddles, known as Griffith's regions that are coupled through long range Josephson coupling and form a vortex glass like phase where the critical field of the rare ordered Griffith's region becomes larger than the upper critical field corresponding to the clean limit QPT. Though the emergence of QGS has mainly been correlated with systems exhibiting SMT due to stronger Josephson coupling with the metallic normal state [15,20,28], of late, its association with SIT based systems has been experimentally demonstrated for a few materials[18,20]. In order to explore the QGS in both type of systems, a wide range of variations in normal state resistance ($R_N$) is needed among the samples. For example, about 20 fold variation in $R_N$ was reported for TiO thin films[20], whereas, 3-4 times variation in resistance was considered for the NbN samples [18]. Here, we demonstrate the observation of QGS in 2D disordered TiN thinfilms with a variation of more than two orders of magnitude in the $R_N$.

Disordered TiN is one of the initial materials which has been able to successfully demonstrate the experimental observation of quantum criticality[2], field- and disorder-induced SIT [2,29,30], multiple crossing points[10] among the other interesting quantum phenomena[2,31-33]. But, till now, no report has shown the existence of QGS in TiN thin films. Here, we report the observation of multiple crossing points in the magnetoresistance isotherms along with a diverging dynamical critical exponent at low temperature while approaching the QPT in disordered TiN thin films of thickness ranging 2-4 nm. A direct activated dynamical scaling analysis on the field and temperature dependent sheet resistance $R_s(B,T)$ for a wide range of temperature further confirms the infinite-randomness fixed critical point to be governing the QPT. Based on the detailed analysis, the emergence of the QGS in TiN thin films is established and the corresponding field-temperature phase diagram features a broad QGS region extending up to a temperature which is very close to the $T_c$ of the sample.

## Results

The temperature dependent sheet resistance $R_s(T)$ measurements for five representative TiN samples of thickness 2-4 nm are presented in Fig. 1. Three samples measured down to 2 K are shown in Fig. 1(a) and the rest two measured down to 300 mK take place in Fig. 1(b). With reduction in thickness, the normal state resistance ($R_N$) increases and the transition temperature ($T_c$) decreases. Here, $T_c$ corresponds to the temperature at which the derivative $dR_s/dT$ becomes maximum. The $T_c$ values closely match with their respective mean field transition temperature $T_{c0}$ as obtained from the quantum corrections to the conductivity (QCC) theories [31,34]. Further, while cooling down from room temperature, an upturn with a resistance maximum $R_{max}$ appears for all the samples just before the superconducting onset indicating a weakly localized metallic state [34]. The appearance of the resistance peak $R_{max}$ at the temperature $T_{max}$ is shown by the arrows in Fig. 1(a) for the sample TN10A and in Supplementary Fig. 1 for samples TN9 & TN8. It is evident that with reduction in thickness, samples become more disordered and more localized as indicated by the increased $R_N$ and the steeper slope for the negative $dR_s/dT$ region of the resistance upturn, respectively.

In the inset of Fig. 1(b), the $R_N$ values for all the samples presented in this article and in Supplementary Fig. 2 (Supplementary Note 1) are summarized with respect to the quantum resistance $R_Q$ which is denoted by the horizontal dashed line separating the two shaded regions in cyan and purple. As for the reference of $R_N$, the resistance $R_{max}$ is plotted against the room temperature resistance $R_{300 K}$. Here, the selected samples offer more than two orders of change in the $R_N$ from ~140 Ω to ~18 kΩ. The wide variations in $R_N$ are obtained by tuning the thickness and the annealing temperature ($T_a$) during the growth process. Two different temperatures for $T_a$, 780 °C and 820 °C, are considered here, and the resistance values for the corresponding samples are shown in olive-green and blue rectangles, respectively. Comparing the resistance with respect to $R_Q$, the samples with resistance less than 500 Ω are considered here as weakly disordered ($R_S<<R_Q$), the samples with resistance more than 1.5 kΩ are considered as moderately disordered ($R_S<R_Q$) and the samples with resistance much higher than $R_Q$ are considered as strongly disordered ($R_S>>R_Q$).

**Emergence of multiple crossing points in magnetoresistance isotherms**

The magnetic-field and temperature dependent sheet resistance $R_s(B,T)$ measurements for the samples measured down to 2K are shown in Fig. 2 where the left panel represents the $R_s(T)$ measured under different field and the right panel displays the magnetoresistance isotherms $R_s(B)$. The sample thickness varies in descending order from the top to the bottom. For comparison among the samples, the variations of $R_s(T)$ with field are placed in equivalent scaling as evident by their normalized resistance $R_s/R_{max}$. With increasing field, a resistance plateau indicating a field induced SMT (or weakly localized metallic transition) [18] is observed for the 4 nm (TN8) and 3 nm (TN9) thick samples, whereas, a relatively strong upturn indicating a SIT or superconductor to weak insulator transition [20] appears for the thinnest sample TN10A (2 nm). Further, a selective region of the measured $R_s(B)$ isotherms from each of the samples TN8, TN9 & TN10A is highlighted in Fig. 2 (b), (d) & (f), respectively. In contrast to the conventional single point crossing, the $R_s(B)$ isotherms, measured between 2 K to 5 K with 0.1 K interval, display a series of crossing points (shown in black filled diamonds) that form a continuous line of SMT/SIT critical points. The appearance of a continuum of MCPs hints towards the existence of QGS at very low temperature ($T \rightarrow 0$) where the critical exponents product $z\nu$ diverges [15,16,21]. Furthermore, a quantitative analysis on the relative span $\Delta B_c/\Delta T$ covered by the MCPs is done with respect to the sample thickness. Here, $\Delta B_c$ is the extent in magnetic field for a specific range of temperature $\Delta T$. The values of $\Delta B_c/\Delta T$ for samples TN8 (4 nm), TN9 (3 nm) and TN10A (2 nm) are found as 0.306 T/K, 0.446 T/K and 0.896 T/K, respectively. Thus, thickness reduction leads to an enhancement in the relative span of the MCPs and from the trend, one can expect a single crossing point in relatively thicker samples. A similar trend is observed in Supplementary Fig. 3 (Supplementary Note 2) for the samples TN4 (4 nm) and TN5A (3 nm) that are annealed at 820 °C.

Fig. 2 indicates that the field extent covered by MCPs increases with increasing disorder and hence the randomness in the critical behavior gets stronger with increasing disorder. In Fig. 3, we present $R_s(B,T)$ measurements for a highly disordered sample (TN6) with $R_N >> R_Q$. The zero-field $R_s(T)$ measured from room temperature down to 2 K as shown in Fig. 3(a) clearly shows a non-metallic behavior in the normal state above the $T_{max}$ with negative $dR_S/dT$ in the whole temperature range from 300 K down to $T_{max}$. Here, the upward slope gets much steeper just above the $T_{max}$ while approaching the $T_{max}$ from higher temperature side and a resistance peak $R_{max}$ appears at the $T_{max}$. This leads to a much higher $R_{max}$ than the resistance $R_{300 K}$ measured at 300 K. Further, for $T<T_{max}$, a resistance drop of about 25% of the $R_{max}$ is observed down to 2 K. The evolution of the transition region of the $R_s(T)$ under magnetic field as shown in the inset of Fig. 3(a) presents a transition from superconducting to insulating regime upon the field variation from 2T to 4T. Further, Fig. 3(b) presents $R_s(B)$ isotherms for the temperature window of $T_{max}$ to 2 K where a continuum of MCPs (shown by the black filled diamonds) is clearly evident. The relative span ($\Delta B_c/\Delta T$) of the MCPs appears as ~ 2.37 T/K which is indeed much higher than the samples shown in Fig. 2 with relatively low $R_N$. Therefore, MCPs are observed in ultrathin TiN ($\leq$ 4 nm) samples spanning in wide range of disorder which indicates the existence of MCPs in both types of field induced QPTs SMT and SIT.

*Emergence of quantum Griffiths singularity*

The continuum of MCPs observed in $R_s(B,T)$ measured down to 2 K (shown in Fig. 2 & 3) indicates towards the existence of QGS in these samples at zero temperature limit [19] where the observation of diverging critical exponent $z\nu$ provides the solid evidence of QGS [28]. Consequently, two samples TN9A (~3 nm) and TN10B (~2 nm) exhibiting field-induced SMT & SIT, respectively, are measured down to 300 mK. The $R_s(B,T)$ measurements for TN9A are presented in Fig. 4. With increasing magnetic field from 3.5 T to 3.75 T in the field dependent $R_s(T)$ as shown in Fig. 4(a), transition from superconducting to weakly localized metallic state is observed at the lowest measurement temperature. Further, Fig. 4(b) presents the $R_s(B)$ isotherms measured in the temperature interval 0.3 K $\leq T \leq$ 3 K where the neighboring $R_s(B)$ isotherms cross at different critical fields (shown in black diamonds) that form a continuous line of MCPs indicating the existence of QGS [19,28]. This continuum of MCPs spans over a wide range of temperature and magnetic field. The temperature dependence of the crossing field $B_c(T)$ is shown in the inset of Fig. 4(b). Here, the average temperature is considered at each crossing point and the error bars represent the range.

The existence of QGS can be established by the divergence of critical exponent $z\nu$ upon approaching the QPT at $T \to 0$ and $B \to B_c^*$, where $B_c^*$ is the characteristic critical field [15,16,21]. Here, the critical exponent $z\nu$ is obtained by adopting the finite size power law scaling (FSS) analysis as is generally used for a continuous QPT associated with a fixed single QCP [7]. Under the power law scaling, the temperature and magnetic field dependent sheet resistance $R_s(B,T)$ near a field induced QPT can be described as [7],

$$R_s(B,T) = R_s^c f(\delta T^{-1/z\nu}) \qquad (1)$$

where, $\delta = |B - B_c|$ is the distance from the critical field $B_c$, $R_s^c$ is the sheet resistance at the critical point and $f(x)$ is the scaling factor with $f(0) = 1$. $z$ and $\nu$ are the dynamical and corelation length critical exponents, respectively. To perform the FSS analysis on the $R_s(B,T)$ data shown in Fig. 4(b), a set of three adjacent $R_s(B)$ isotherms crossing at a particular critical field $B_c$ in each small temperature interval is considered for extracting $z\nu$ values by using the power law scaling in Eq. (1). The detailed FSS analysis is illustrated in Supplementary Fig. (4)-(6) in the Supplementary Note 3. The extracted $z\nu$ values are plotted against the critical fields in Fig. 4(c). Initially, $z\nu$ increases slowly with magnetic field from 3 K down to ~ 1.55 K, with further lowering temperature, $z\nu$ increases rapidly with field and shows a diverging behavior while approaching the characteristic critical field $B_c^*$. This reflects an activated scaling behavior of $z\nu$ with respect to the field near the characteristic critical field $B_c^*$ and is expressed as [35],

$$z\nu \approx C(B - B_c^*)^{-\nu\psi}, \qquad (2)$$

where $C$ is a constant and $\nu = 1.2$ & $\psi = 0.5$ are the corelation length exponent and the tunneling exponent associated with the 2D infinite-randomness fixed critical point (IRFCP), respectively [35-37]. The activated scaling law is in good agreement with the experimental data as shown by the red solid curve in Fig. 4(c) which yields $B_c^*$=3.78 T. This indicates towards the existence of IRFCP justifying the presence of QGS in the sample.

Further, a QPT governed by an IRFCP can be described by a direct activated dynamical scaling as a whole where the infinite-randomness critical exponent $\nu\psi$ can replace the critical exponent $z\nu$ associated with the conventional power law scaling. Here, at $T \rightarrow 0$ the critical exponent products from power law scaling ($z\nu$) and activated dynamical scaling ($\nu\psi$) are related as [28],

$$\left(\frac{1}{zv}\right) = \left(\frac{1}{v\psi}\right) \cdot \frac{1}{ln(T_0/T)} \tag{3}$$

where the exponent $v\psi$ ( $v$ is the corerelation length critical exponent and $\psi$ is the tunneling exponent) represents the universality class of the QPT associated with IRFCP. The fit using Eq. (3) to the temperature dependent $zv$ (shown in Supplementary Fig. 7 in the Supplementary Note 4) leads to the characteristic temperature $T_0$ and the critical exponent $v\psi$ which are used as free parameters. Further, Eq. (3) indiates a diverging $zv$ while approaching to $T \to 0$, confirming the QCP is of IRFCP type where physical quantities can be described by the direct activated dynamical scaling. Consequently, the sheet resistance $R_s$ can be expressed as [28],

$$R_s\left(\delta, ln\frac{T_0}{T}\right) = \Phi\left[\delta\left(ln\frac{T_0}{T}\right)^{1/v\psi}\right] \tag{4}$$

where, $\delta = |B_c - B_c^*|/B_c^*$ is the distance from the critical field $B_c^*$ associated to the IRFCP and $\Phi$ corresponds to the scaling function. Here, $T_0$ and $v\psi$ are the same as have been used in Eq. (3). Further, the crossing field $B_c(T)$ near the transition is also expected to follow the activated dynamical scaling with irrelevant scaling correction as [28],

$$B_c(T) = B_c^* \times \left[1 - u\left(ln\frac{T_0}{T}\right)^{-p}\right] \tag{5}$$

with the exponent $p = \frac{1}{v\psi} + \frac{\omega}{\psi}$ and $u$ is the leading irrelavant scaling variable responsible for the correction to the scaling. $\omega$ is the associated irrelevant exponent which is always positive [22]. The red solid curve in the inset of Fig. 4(b) represents the fit to the $B_c(T)$ phase boundary using Eq. (5) where $B_c^*$, $u$ and $p$ are considered as the fitting parameters. $T_0$= 4.41 K and $v\psi = 0.63$ are obtained from the fit using Eq. (3) [Supplementary Fig. 7(a) in the Supplementary Note 4]. Here, $v\psi = 0.63$ closely matches with the theoretically predicted value of ~ 0.6 with $v \sim 1.2$ & $\psi \sim 0.5$ [35-37]. The best fit values for $B_c^*$, $u$ and $p$ are obtained as (3.77 ± 0.02) T, (0.115 ± 0.007) and (1.92 ± 0.15), respectively.

Further, in Fig. 4(d), the direct activated dynamical scaling using Eq. (4) has been applied on a set of $R_s(B)$ isotherms and the best collapse is obtained for $B_c^* = 3.77$ T, $T_0 = 4.56 \pm 0.23$ K and $v\psi = 0.63$ in the temperature interval 0.3 K $\leq T \leq$ 1.75 K. Here, $B_c^*$ is obtained by minimizing the variance in magnetoresistance isotherms when plotted against the scaling parameter [18,28]. The same value for $B_c^*$ is obtained from the phase boundary fitting of $B_c(T)$ using Eq. (5) [inset of Fig. 4(b)]. Furthermore, $B_c^* =$ 3.77 T, agrees closely to 3.78 T which is obtained from the fit to the field dependence of $z\nu$ using Eq. (2) [shown in Fig. 4(b)]. The characteristic temperature $T_0$, used as an adjusting parameter to achieve the best collapse, is obtained as 4.56 ± 0.23 K which is close to the temperature 4.41 K as found from the fit of $z\nu$ versus temperature using Eq. (3) (Supplementary Fig. 7 in the Supplementary Note 4). It is observed in Fig. 4(d) that for large values of the scaling parameter away from $B_c^*$, the scaling collapse starts to break down at ~ 4.29 T and ~1.85 T, for the upper and lower branches, respectively. These breakdowns can be attributted to the negligible quantum fluctuations occuring in the normal state at high magnetic field (upper branch) and in the superconducting state at very low magnetic field and low temperature (lower branch) [18,28]. It should be noted that the temperature window of 0.3 K ≤ $T$ ≤ 1.75 K remains the same for (i) the direct activated scaling of the MR isotherms using Eq. (4) [in Fig. 4(d)], (ii) the $B_c(T)$ phase boundary fitting using Eq. (5) [in the inset of Fig. 4(b)] and (iii) the $z\nu$ vs. temperature fit using Eq. (3) [in Supplementary Fig. 7 in the Supplementary Note 4]. However, at high temperature, the quantum fluctuation becomes insignificant compared to the thermal fluctuations and the data set starts to deviate from the activated scaling.

Next in Fig. 5, we present the $R_s(B,T)$ measurements carried out at temperature down to 300 mK for a more resistive ($R_N$~ 2 kΩ) sample TN10B (~ 2 nm) [shown in Fig 1(b)] with weak insulating background where the formation of rare ordered superconducting puddles (Griffiths regions) via long-range Josephson coupling becomes more challenging compared to a metallic background. A selected region from the set of $R_s(T)$, measured under fixed perpendicular fields from 0 to 5 T in 0.25 T steps, is shown in Fig. 5(a). The trace of superconductivity with positive downward slope is observed for field up to 3.75 T. But, a clear upward transition with negative slope (d$R_s$/d$T$ < 0 ) occurs for an increased field of 4 T. Here, as the $R_N$ is lower than the R$_Q$, the logarithmic derivative of the sheet conductance ($\sigma_s = 1/R_s$) is considered for examining the state for $B \geq 4$ T. Though d$R_s$/d$T$ is negative but often this does not ascertain whether a sample is metallic or insulating [18,38]. It is shown that the logarithmic derivative of sheet conductance $w = \mathrm{dln}\sigma_s/\mathrm{dln}T$ at zero-temperature limit ($T \to 0$) is more sensitive than the derivative $dR_s/dT$ in determining whether a sample is metallic or insulating. For $w > 0$, a positive value of $dw/dT$ indicates a metallic state whereas, a negative value of $dw/dT$ indicates the sample to be very likely insulating [39]. In the inset of Fig. 5(a), $w = \mathrm{dln}\sigma_s/\mathrm{dln}T$ corresponding to the $R_s(T)$ measured under 4 T is plotted against $T^{1/2}$. Here, $w$ is positive and towards the lowest measurement temperature $w$ increases with decreasing temperature leading to a negative value of $dw/dT$. This indicates a field induced superconducting to insulating quantum phase transition (SIT) at zero-temperature limit for $B \geq 4$ T. The insulating phase near the transition has been investigated further by $R_s(B)$ isotherms measured at low temperature down to 300 mK and for a magnetic field up to 10 T. The corresponding magnetoresistance (MR) data is shown in Supplementary Fig. 10(b) in the Supplementary Note 6 which unveils a broad resistance peak for $B \geq 4$ T. The appearance of resistance peak near the transition indicates the field-induced localization of Cooper pairs [2,29,40,41]. However, as the $R_N$ is much lower than the quantum resistance $R_Q$ of Cooper pairs, a parallel contribution from fermionic channel of quasiparticles is most likely to be present in addition to the bosonic contribution from localized Cooper pairs. Further, the logarithmic temperature dependence of the conductance [Supplementary Fig. 10(c-d) in the Supplementary Note 6] at higher field confirms the fermionic scenario [5,34,42,43].

Further, a detailed view near the field induced transition to the insulating state is shown in Fig. 5(b) by the $R_s(B)$ isotherms measured in the temperature interval 0.3 K $\leq T \leq$ 3.0 K. A continuum of MCPs (shown by the black diamonds) is evident. The phase boundary as represented by the temperature dependent crossing field $B_c(T)$ is displyed in the inset of Fig 5(b). The red solid curve represents the activated dynamical fit using Eq. (5) for the temperature range 0.3 K $\leq T \leq$ 1.75 K with $T_0$= 3.11 K and $v\psi = 0.62$ as obtained from the fit using Eq. (3) to the temperature dependent $zv$ (Supplementary Fig. 7(b) in the Supplementary Note 4). The critical field $B_c^*$ is obtained as 4.03 ± 0.02 T. The best fit values for the parameters $u$, $p$ and the critical field $B_c^*$ are obtained as 0.031 ± 0.006, 2.15 ± 0.26 and 4.03 ± 0.02 T, respectively.

The field dependent $z\nu$ is plotted in Fig. 5(c) which indicates a diverging $z\nu$ while approaching the characteristic critical field $B_c^*$ at the low temperature limit. The experimental data is in good agreement with the activated scaling law as described by the red solid curve using Eq. (2) indicating towards the existence of IRFCP transition which justifies the presence of QGS in the systems. The best fit value for the characteristic field $B_c^*$ is obtained as 4.03 T. Further in Fig. 5(d), a direct activated dynamical scaling using Eq. (4) is performed on the $R_s(B,T)$. Here, the best collapse is observed for $B_c^* = 4.0$ T in the temperature range 0.3 K ≤ $T$ ≤ 1.75 K with $\nu\psi$ fixed at the value of 0.62. The best collapse, obtained by using the characteristic temperature $T_0$ as an adjusting parameter, leads to $T_0$ = 3.27± 0.25 K. The critical field $B_c^* = 4.0$ T is very close to the value of 4.03 T which is obtained from the phase boundary fitting of $B_c(T)$ [inset of Fig 5(b)] as well as from the activated scaling law as presented in Fig. 5(c). Further, the characteristic temperature $T_0$ is in good agreement too with the value obtained from the temperature dependence of $z\nu$ shown in Supplementary Fig. 7(b) in the Supplementary Note 4. Here, the scaling collapse breaks down at high temperature and at fields far from the critical field $B_c^*$ due to the diminishing quantum fluctuations [28].

Here, it is noteworthy to mention that Eq. (4) associated with the direct activated dynamical scaling of the MR isotherms predicts a single critical point $B_c^*$ which does not account for the temperature dependence of the crossing fields as observed by the MCPs. The scaling behaviors, shown in Fig. 4(d) and in Fig. 5(d), confirm the existence of a single critical point in the magnetic field for a wide range of temperature from 300 mK to about 1.75 K for both the samples TN9A and TN10B. Further, Eq. (3) indicates a diverging behaviour of the critical exponent $z\nu$ with temperature at $T \to 0$ and the corresponding fit shown in Supplementary Fig. 7 establishes the diverging behavior of $z\nu$ at low temperature. These observations confirm the QPT to be governed by an IRFCP. However, there is a clear difference in the temperature dependent critical field $B_c(T)$ for these two samples TN9A (SMT category) and TN10B (SIT category). For the former, the critical field moves towards upward direction while approaching the zero-temperature limit [inset of Fig. 4(b)], wheras, for the latter, $B_c$ increases initially with decreasing temperature in the high temperature regime ($T \gtrsim 1$ K) but finally it tends to saturate around $B_c^*$ (4.03 T) as the temperature approaches to 0 K [inset of Fig. 5(b)]. The upward trend of $B_c(T)$ as observed for TN9A is similar to the other reported SMT systems exihibiting QGS [15,18,20,24-26], whereas, a saturating $B_c$ at zero-temperature limit is one of the features observed in systems showing SIT [20]. The differences in the behaviour of $B_c$ at $T \to 0$ might be related to the different Josephson interaction strength among the superconducting puddles with insulating & metallic normal states, for the SIT and SMT systems, respectively.

## Discussion

Based on the low temperature magnetotransport measurements, a comprehensive *B-T* Phase diagram has been constructed for the 2-nm thick sample TN10B for which the normal state is weakly insulating and is considered in the category of SIT system. The phase diagram is shown in Fig. 6 where $T_c^{Onset}(B)$ (the half-filled green diamonds) and the crossing field $B_c(T)$ (the pink squares) follow the same track. Here, the $T_c^{Onset}(B)$ separates the weakly localized insulating regime with negative $dR_s/dT$ from the superconducting fluctuation regime with positive $dR_s/dT$ which is mainly driven by the thermal fluctuation at the high temperature limit for $T > T_m = 1.75$ K. The separation of the weakly localized insulating normal state from the thermal fluctuation regime is shown by the blue broken curve, connecting the $T_c^{Onset}(B)$ and the $B_c(T)$ above 1.75 K. The mean field critical field $B_c^{MF}(T)$ (shown by the blue spheres) is obtained from the field dependent $R_s(T)$ by using the Ullah-Dorsay (UD) scaling method [21,44-46] by which in actuality, the mean field critical temperature $T_c^{MF}(B)$ is extracted for a particular field. The UD scaling is done for the field range 1.5 T ≤ B ≤ 3.25 T and outside of this window the scaling deviates (The details are shown in Supplementary Fig. 9 in the Supplementary Note 5). It is found that the obtained $T_c^{MF}$ closely matches with the $T_c$. Accordingly, for field lower than 1.5 T, $T_c^{MF}(B)$ is represented by the $T_c(B)$. A fit using the empirical formula [16,18] $B_c^{MF}(T) = B_c^{MF}(0)[1 - (T/T_c)^2]$ appears to be in good agreement with the mean field critical field $B_c^{MF}(T)$ as represented by the solid back curve. From the fit, the upper critical field $B_c^{MF}(0)$ and the $T_c(0)$ are obtained as 3.65 T and 1.9 K, respectively. At temperature lower than the mean field critical temperature $T_c^{MF}(B)$, phase fluctuation is the dominant mechanism which relates to the thermally activated flux flow (TAFF) at high temperature and to quantum creep at low temperature [21]. These two regions are separated by the characteristic temperature $T_{TAFF}$ as represented by the purple triangles. $T_{TAFF}$ is extracted from the Arrhenius plot analysis of the logarithmic of $R_s(T)$ measured under a particular field when plotted against inverse temperature ($T^{-1}$). The $T_{TAFF}$ represents the crossover temperature between the TAFF regime (pink region) and the quantum phase fluctuation regime (blue region). The quantum phase fluctuation regime below the $T_{TAFF}(B)$ is marked as the condensed superconducting (SC) phase.

Finally, the QGS region is marked based on the activated dynamical scaling analysis associated with the IRFCP as presented in Fig. 5. Here, the magnetoresistance data can be adequately described by the activated dynamical scaling [Eq. (4)] for the temperature span of 0.3 K ≤ $T$ ≤ 1.75 K. In the same temperature window, the $B_c(T)$ can be portrayed by the IRFCP governed activated dynamical scaling with the irrelevant scaling corrections [Eq. (5)] as shown by the green solid curve in Fig. 6. Thus, the activated dynamical scaling analysis on the measured data suggests for a wide range of temperature up to 1.75 K above the $B_c^{MF}(0)$ as the region of QGS. This is consistent with the field dependence of $z\nu$ presented in Fig. 5(c) which displays a plateau region where $z\nu$ remains almost constant with magnitude < 1 for $B \lesssim$ 3.8 T and $T \gtrsim$ 1.75 K. For temperature below 1.75 K, $z\nu$ starts to rise sharply with the field and finally it diverges while approaching the critical field $B_c^*$. The similar trend is observed also in the literature where, initially at high temperature, $z\nu$ shows a plateau region ($z\nu$ ~ 0.5 <1) with increasing magnetic field, but with decreasing temperature, $z\nu$ changes very rapidly with the field and diverges while approaching the characteristic critical field $B_c^*$ associated with the IRFCP transition, thus, establishing the presence of QGS [15,17,19-21,24,25]. Accordingly, the region extending in temperature from absolute zero to 1.75 K above the mean field upper critical field $B_c^{MF}(0)$ can be denoted as the region of QGS which is highlighted by the light orange shade in Fig. 6. In this region, field variation is very weak (as it tends to saturate towards zero temperature) compared to a relatively wide variation in temperature. At the higher temperature side within the QGS region, though the exponent $z\nu$ increases with the field but it remains less than unity ($z\nu < 1$). Whereas, close to the critical field $B_c^*$ for T ≤ 0.6 K in the QGS region, a diverging behavior of $z\nu$ with field is observed and the condition $z\nu > 1$ is achieved. Accordingly, the quantum phase fluctuation regime below the $T_{TAFF}$ (B) is extended up to the critical field $B_c^*$ by the purple dashed curve which separates the QGS region with $z\nu > 1$ from that with $z\nu < 1$.

In the QGS regime, the dominant quenched disorders lead to the formation of superconducting puddles in which the superconducting order parameter locally gets concentrated and transition to a stable macroscopic phase-coherent state can be achieved when the Josephson coupling between these rare ordered puddles becomes strong enough [47]. The coherent coupling between the puddles depends on the characteristic correlation length through which the phase coherence is preserved [9]. The correlation length varies inversely with the temperature for long range coupling [7]. Near QPT, the correlation length is smaller than the distance between the puddles but comparable to the size of the individual puddle and the transport is dominated by the quantum fluctuation of the superconducting order parameter and can be considered as the region of quantum fluctuation [47]. The existence of QGS at relatively high temperature is not commonly observed due to the smearing out of the effect of the quenched disorder by the thermal fluctuations and the rare ordered regions of superconducting puddles may face survival issues against the thermal fluctuations [15]. However, in the present study as determined by the direct activated scaling, a reasonably large portion in the phase diagram is designated as the QGS regime where the region with $z\nu > 1$ represents the absolute dominance of QGS with diverging critical exponents and the portion with $z\nu < 1$ extending up to 1.75 K indicates the onset of QGS with a significant influence from the thermal and quantum fluctuations of the order parameter. The wider span in temperature has been reported recently for nonepitaxial NbN [19]. Here, the appearance of QGS for a wide temperature range and with a large variation in the $R_N$ strongly indicate towards the robustness of QGS in these nonepitaxial polycrystalline TiN thin film samples that are inherently disordered by the presence of other non-superconducting elements and phases [48].

Theoretically, the QGS associated with SMT/SIT has been studied by the similar superconducting droplet model where dilute but large droplets/puddles of phase-coherent rare ordered regions are separated by the disordered metallic/insulating background acting as the dissipative medium which helps to slow down the dynamics of the rare ordered regions. Here, the model assumes an isolated droplet picture where the distance between the neighbouring droplets is more than the superconducting correlation length so that no percolating path is available. In order for QGS to emerge, the susceptibility of an isolated droplet should diverge while approaching the QCP. Here, the probability $P_{droplet}$ for the formation of large superconducting droplets falls exponentially with its size as, $P_{droplet} \sim exp(-bL^d)$ whereas, the susceptibility of the individual droplet $\chi_{droplet}$ grows exponentially with the size as, $\chi_{droplet} \sim exp(aL^d)$. $L$ is the lateral dimension of the droplet and $d$ refers to the space dimensionality. $b$ relates with the disorder strength and $c$ represents the distance from the criticality [47]. Therefore, the existence of QGS depends strongly on the balance between $P_{droplet}$ and $\chi_{droplet}$ [49]. Further, the dissipation, caused by the gapless electrons that can penetrate the superconducting droplets entirely, can change the afore-mentioned balance dramatically. The gapless electrons in the dissipative bath can penetrate these large droplets only up to a maximum distance of the coherence length. As it is argued [47,50], for the droplets that are large compared to the coherence length, the coupling to the dissipative bath grows at most in proportion to the surface length in 2D (surface area in 3D) and hence, the susceptibility varies exponentially with the perimeter instead of area in 2D (the surface area rather than the volume in 3D). Thus, the critical dimensionality of the rare region is below the lower critical dimension of the problem, which leads to the conventional QPT [51].

Here, the main issue relates with two length scales that are size and the coherence length of the puddles. It has been shown that the energy scale related to the rare-ordered region falls exponentially with the size of the droplet which leads to the singular density of states for anomalously large droplets[52,53]. Hence, the dimensional crossover for anomalously large droplets is expected to occur at extremely low temperature[28,49] where the characteristic correlation length becomes larger than the separation between the superconducting puddles and the transition can still be represented by a particular universality class as described by appropriate critical exponents in the case of conventional QPT [47]. Whereas, for the case of QGS, there exists an optimal size which is of the order of coherence length with an optimal range for the radius of the droplets. Above this optimal range, the probability of finding a rare-ordered droplet becomes extraordinarily rare and below this range, the susceptibility of the droplets becomes small [47]. Hence, QGS can be expected for the droplets having the dimensions falling within this optimal range which extends up to the limit where the droplets still follow the quantum dynamics and the QGS phase arises from the quantum fluctuations of the superconducting order parameter in the droplets [49,52].

Now, according to the BCS theory, the coherence length $\xi_{BCS} = \frac{\hbar v_F}{\pi \Delta(0)}$, becomes anomalously large while approaching the QCP where superconducting energy gap $\Delta(0) \to 0$. Here, $v_F$ is the Fermi velocity. Further, from the fluctuation spectroscopy studied in 2D superconductor under perpendicular magnetic field, the existence of quantum liquid phase with long coherence length for the field range $B_{c2} < B < B_c^*$ is expected due to the quantum fluctuations among the coherently rotating fluctuating Cooper pairs (FCPs) that form the rare-ordered superconducting droplets [54,55]. In this field region, the quantum fluctuation driven spatial coherence length $\xi_{QF}$ becomes much longer than $\xi_{BCS}$ ($\xi_{QF} \gg \xi_{BCS}$) and $\xi_{QF}$ increases with increasing field above the $B_{c2}$ till it reaches to the QCP. Here, the coherence length $\xi_{QF}$ measures the characteristic size of the superconducting droplets [55]. Therefore, the assumption that the gapless electron bath causing the dissipation can penetrate into the droplet is self-consistent in this region of field near the QCP where the coherence length is much larger due to the quantum fluctuations.

## Conclusions

In conclusion, a systematic investigation through low temperature magnetotransport measurements in disordered TiN thinfilms of thickness in the range of 2-4 nm demonstrates field induced superconductor to metal (SMT) and/or superconductor to (weak) insulator (SIT) type QPTs that are associated with multiple crossing points (MCPs) in magnetoresistance isotherms for a number of samples varying more than two orders of magnitude in their $R_N$ values. Measurements down to 300 mK manifest a diverging behavior of the dynamical critical exponent $zv$ upon approaching the QPT indicating the emergence of QGS during the transition. Further, the applicability of the activated dynamical scaling to describe the magnetoresistance isotherms for a wide range of temperature suggests the universality class of the transition which is governed by the infinite-randomness fixed critical point. The association of QGS with SMT and SIT can be distinguished by the appearance of an upward trend of $B_c(T)$ and a saturating $B_c(T)$, respectively. Finally, a comprehensive phase diagram is established for the SIT system which displays a wide QGS region. The robustness of the QGS in the presented nonepitaxial TiN thin films is confirmed by its existence over a wide temperature range and also over a wide variation in the disorder strength as evident in their $R_N$ values.

## Methods

We have employed an intrinsic Si (100) substrate covered with $Si_3N_4$ dielectric spacer layer of 80 nm thickness for the thin films samples to be prepared on. The $Si_3N_4$ topping layer was grown by using low pressure chemical vapor deposition (LPCVD) technique and in this study, $Si_3N_4$ is the only source of nitrogen for the nitridation of Ti to produce TiN thin films [56]. After adopting a standard cleaning process for the $Si_3N_4$/Si (100) substrates, Ti films were deposited on the substrate by dc magnetron sputtering using a Ti target of 99.995% purity in the presence of high purity Ar (99.9999%) gas. Sputtering of Ti was performed with a base pressure less than 1.5 x $10^{-7}$ Torr. Subsequently, the sputtered Ti films were transferred *in situ* to an attached UHV chamber for annealing. Different batches were prepared by varying the annealing temperature ($T_a$) and the thickness of the Ti films. The annealing temperatures used to prepare the samples presented in this study were 780 °C and 820 °C with a variation of ± 10 °C. The annealing was done for 2 hours at a pressure less than 5 x $10^{-8}$ Torr. During the annealing process, $Si_3N_4$ substrate decomposed into Si (s) and N (g) atoms and due to high reactive nature of Ti towards both Si and N, the annealing process led to the formation of superconducting TiN as a majority phase along with metallic $TiSi_2$ as a minority phase. The detailed chemical kinetics involved in the substrate mediated nitridation technique is reported elsewhere [48]. For low temperature transport measurements, thin films in multi-terminal device geometry were grown by using shadow mask made of stainless-steel. Further, electrical contact leads of Au (80-100 nm)/Ti (5 nm) were deposited by dc magnetron sputtering by using complimentary shadow masks.

Low temperature resistivity measurements from room temperature down to 2 K were carried out in a Physical Property Measurement System (PPMS) equipped with a 16 T magnet from Quantum Design at UGC-DAE CSR Indore. The same measurement system was used to measure down to 300 mK with the help of a dilution refrigerator insert/port/unit. The electrical resistivity were measured with an excitation of 100 nA.

## Data availability

The data that represent the results in this paper and the data that support the findings of this study are available from the corresponding author upon reasonable request.

## References

1       Dubi, Y., Meir, Y. & Avishai, Y. Nature of the superconductor–insulator transition in disordered superconductors. *Nature* **449**, 876-880 (2007).

2       Baturina, T. I., Strunk, C., Baklanov, M. R. & Satta, A. Quantum Metallicity on the High-Field Side of the Superconductor-Insulator Transition. *Physical Review Letters* **98**, 127003 (2007).

3       Del Maestro, A., Rosenow, B., Hoyos, J. A. & Vojta, T. Dynamical Conductivity at the Dirty Superconductor-Metal Quantum Phase Transition. *Physical Review Letters* **105**, 145702 (2010).

4       Marković, N., Christiansen, C. & Goldman, A. M. Thickness--Magnetic Field Phase Diagram at the Superconductor-Insulator Transition in 2D. *Physical Review Letters* **81**, 5217-5220 (1998).

5        Yazdani, A. & Kapitulnik, A. Superconducting-Insulating Transition in Two-Dimensional a-MoGe Thin Films. *Physical Review Letters* **74**, 3037-3040 (1995).

6        Wang, T. *et al.* Universal relation between doping content and normal-state resistance in gate voltage tuned ultrathin $Bi_2Sr_2CaCu_2O_{8+x}$ flakes. *Physical Review B* **106**, 104509 (2022).

7        Sondhi, S. L., Girvin, S. M., Carini, J. P. & Shahar, D. Continuous quantum phase transitions. *Reviews of Modern Physics* **69**, 315-333 (1997).

8        Belitz, D., Kirkpatrick, T. R. & Vojta, T. How generic scale invariance influences quantum and classical phase transitions. *Reviews of Modern Physics* **77**, 579-632 (2005).

9        Biscaras, J. *et al.* Multiple quantum criticality in a two-dimensional superconductor. *Nature Materials* **12**, 542-548 (2013).

10       Kronfeldner, K., Baturina, T. I. & Strunk, C. Multiple crossing points and possible quantum criticality in the magnetoresistance of thin TiN films. *Physical Review B* **103**, 184512 (2021).

11       Harris, A. B. Effect of random defects on the critical behaviour of Ising models. *Journal of Physics C: Solid State Physics* **7**, 1671 (1974).

12       Vojta, T. & Hoyos, J. A. Criticality and Quenched Disorder: Harris Criterion Versus Rare Regions. *Physical Review Letters* **112**, 075702 (2014).

13       Vojta, T. & Sknepnek, R. Critical points and quenched disorder: From Harris criterion to rare regions and smearing. *physica status solidi (b)* **241**, 2118-2127 (2004).

14       Fisher, D. S. Phase transitions and singularities in random quantum systems. *Physica A: Statistical Mechanics and its Applications* **263**, 222-233 (1999).

15       Xing, Y. *et al.* Quantum Griffiths singularity of superconductor-metal transition in Ga thin films. *Science* **350**, 542-545 (2015).

16       Zhang, E. *et al.* Signature of quantum Griffiths singularity state in a layered quasi-one-dimensional superconductor. *Nature Communications* **9**, 4656 (2018).

17       Huang, C. *et al.* Observation of thickness-tuned universality class in superconducting β-W thin films. *Science Bulletin* **66**, 1830-1838 (2021).

18    Jing, T.-Y. *et al.* Quantum phase transition in NbN superconducting thin films. *Physical Review B* **107**, 184515 (2023).

19    Wang, X. *et al.* Robust quantum Griffiths singularity above 1.5 K in nitride thin films. *Physical Review B* **107**, 094509 (2023).

20    Zhang, C. *et al.* Quantum Griffiths singularities in TiO superconducting thin films with insulating normal states. *NPG Asia Materials* **11**, 76 (2019).

21    Saito, Y., Nojima, T. & Iwasa, Y. Quantum phase transitions in highly crystalline two-dimensional superconductors. *Nature Communications* **9**, 778 (2018).

22    Liu, Y. *et al.* Observation of In-Plane Quantum Griffiths Singularity in Two-Dimensional Crystalline Superconductors. *Physical Review Letters* **127**, 137001 (2021).

23    Xing, Y. *et al.* Ising Superconductivity and Quantum Phase Transition in Macro-Size Monolayer NbSe2. *Nano Letters* **17**, 6802-6807 (2017).

24    Han, X. *et al.* Disorder-Induced Quantum Griffiths Singularity Revealed in an Artificial 2D Superconducting System. *Advanced Science* **7**, 1902849 (2020).

25    Shen, S. *et al.* Observation of quantum Griffiths singularity and ferromagnetism at the superconducting LaAlO_3SrTiO_3(110) interface. *Physical Review B* **94**, 144517 (2016).

26    Zhao, Y. *et al.* Quantum Griffiths Singularity in a Layered Superconducting Organic–Inorganic Hybrid Superlattice. *ACS Materials Letters* **3**, 210-216 (2021).

27    Saito, Y., Nojima, T. & Iwasa, Y. Highly crystalline 2D superconductors. *Nature Reviews Materials* **2**, 16094 (2016).

28    Lewellyn, N. A. *et al.* Infinite-randomness fixed point of the quantum superconductor-metal transitions in amorphous thin films. *Physical Review B* **99**, 054515 (2019).

29    Baturina, T. I. *et al.* Superconductivity on the localization threshold and magnetic-field-tuned superconductor-insulator transition in TiN films. *Journal of Experimental and Theoretical Physics Letters* **79**, 337-341 (2004).

30  Sacépé, B. *et al.* Disorder-Induced Inhomogeneities of the Superconducting State Close to the Superconductor-Insulator Transition. *Physical Review Letters* **101**, 157006 (2008).

31  Sacépé, B. *et al.* Pseudogap in a thin film of a conventional superconductor. *Nature Communications* **1**, 140 (2010).

32  Faley, M. I., Liu, Y. & Dunin-Borkowski, R. E. Titanium Nitride as a New Prospective Material for NanoSQUIDs and Superconducting Nanobridge Electronics. *Nanomaterials* **11** (2021).

33  Yadav, S., Aloysius, R. P., Gupta, G. & Sahoo, S. Granularity mediated multiple reentrances with negative magnetoresistance in disordered TiN thin films. *Scientific Reports* **13**, 22701 (2023).

34  Yadav, S., Kaushik, V., Saravanan, M. P. & Sahoo, S. Probing electron-electron interaction along with superconducting fluctuations in disordered TiN thin films. *Physical Review B* **107**, 014511 (2023).

35  Vojta, T., Farquhar, A. & Mast, J. Infinite-randomness critical point in the two-dimensional disordered contact process. *Physical Review E* **79**, 011111 (2009).

36  Del Maestro, A., Rosenow, B., Müller, M. & Sachdev, S. Infinite Randomness Fixed Point of the Superconductor-Metal Quantum Phase Transition. *Physical Review Letters* **101**, 035701 (2008).

37  Kovács, I. A. & Iglói, F. Renormalization group study of the two-dimensional random transverse-field Ising model. *Physical Review B* **82**, 054437 (2010).

38  Möbius, A. Comment on ``Critical behavior of the zero-temperature conductivity in compensated silicon, Si:(P,B)''. *Physical Review B* **40**, 4194-4195 (1989).

39  Möbius, A. The metal-insulator transition in disordered solids: How theoretical prejudices influence its characterization A critical review of analyses of experimental data. *Critical Reviews in Solid State and Materials Sciences* **44**, 1-55 (2019).

40  Sambandamurthy, G., Engel, L. W., Johansson, A. & Shahar, D. Superconductivity-Related Insulating Behavior. *Physical Review Letters* **92**, 107005 (2004).
23

41  Baturina, T. I., Mironov, A. Y., Vinokur, V. M., Baklanov, M. R. & Strunk, C. Hyperactivated resistance in TiN films on the insulating side of the disorder-driven superconductor-insulator transition. *JETP Letters* **88**, 752-757 (2008).

42  Aubin, H. *et al.* Magnetic-field-induced quantum superconductor-insulator transition in $Nb_{0.15}Si_{0.85}$. *Physical Review B* **73**, 094521 (2006).

43  Zhang, X. *et al.* Size dependent nature of the magnetic-field driven superconductor-to-insulator quantum-phase transitions. *Communications Physics* **4**, 100 (2021).

44  Theunissen, M. H. & Kes, P. H. Resistive transitions of thin film superconductors in a magnetic field. *Physical Review B* **55**, 15183-15190 (1997).

45  Ullah, S. & Dorsey, A. T. Critical fluctuations in high-temperature superconductors and the Ettingshausen effect. *Physical Review Letters* **65**, 2066-2069 (1990).

46  Ullah, S. & Dorsey, A. T. Effect of fluctuations on the transport properties of type-II superconductors in a magnetic field. *Physical Review B* **44**, 262-273 (1991).

47  Spivak, B., Oreto, P. & Kivelson, S. A. Theory of quantum metal to superconductor transitions in highly conducting systems. *Physical Review B* **77**, 214523 (2008).

48  Yadav, S. & Sahoo, S. Interface study of thermally driven chemical kinetics involved in Ti/Si3N4 based metal-substrate assembly by X-ray photoelectron spectroscopy. *Applied Surface Science* **541**, 148465 (2021).

49  Millis, A. J., Morr, D. K. & Schmalian, J. Quantum Griffiths effects in metallic systems. *Physical Review B* **66**, 174433 (2002).

50  Kapitulnik, A., Kivelson, S. A. & Spivak, B. Colloquium: Anomalous metals: Failed superconductors. *Reviews of Modern Physics* **91**, 011002 (2019).

51  Vojta, T. Quantum Griffiths Effects and Smeared Phase Transitions in Metals: Theory and Experiment. *Journal of Low Temperature Physics* **161**, 299-323 (2010).

52  Vojta, T. & Schmalian, J. Quantum Griffiths effects in itinerant Heisenberg magnets. *Physical Review B* **72**, 045438 (2005).

53      Nozadze, D. & Vojta, T. Numerical method for disordered quantum phase transitions in the large-N limit. *physica status solidi (b)* **251**, 675-682 (2014).

54      Glatz, A., Varlamov, A. A. & Vinokur, V. M. Fluctuation spectroscopy of disordered two-dimensional superconductors. *Physical Review B* **84**, 104510 (2011).

55      Varlamov, A. A., Galda, A. & Glatz, A. Fluctuation spectroscopy: From Rayleigh-Jeans waves to Abrikosov vortex clusters. *Reviews of Modern Physics* **90**, 015009 (2018).

56      Yadav, S. *et al.* Substrate Mediated Synthesis of Ti–Si–N Nano-and-Micro Structures for Optoelectronic Applications. *Advanced Engineering Materials* **21**, 1900061 (2019).

## Acknowledgements

We highly acknowledge UGC-DAE CSR, Indore, India for carrying out the low temperature resistivity measurements in PPMS. S.Y. acknowledges the Senior Research fellowship (SRF) from UGC. This work was supported by CSIR network project 'AQuaRIUS' (Project No. PSC 0110) and is carried out under the mission mode project 'SAMARTH' (Project No. HCP-55) on 'Quantum Current Metrology'.

## Author contributions

S.Y. and S.S. fabricated the samples. S.Y. and M.P.S. carried out the low temperature transport measurements. S.Y. and S.S. analyzed the data and wrote the manuscript. S.S. planned, designed and supervised the study. All the authors have read and reviewed the manuscript.

**Competing interests:** The authors declare no competing financial and/or non-financial interests in relation to the work described.

## Additional information

**Supplementary Information** accompanies this paper at.

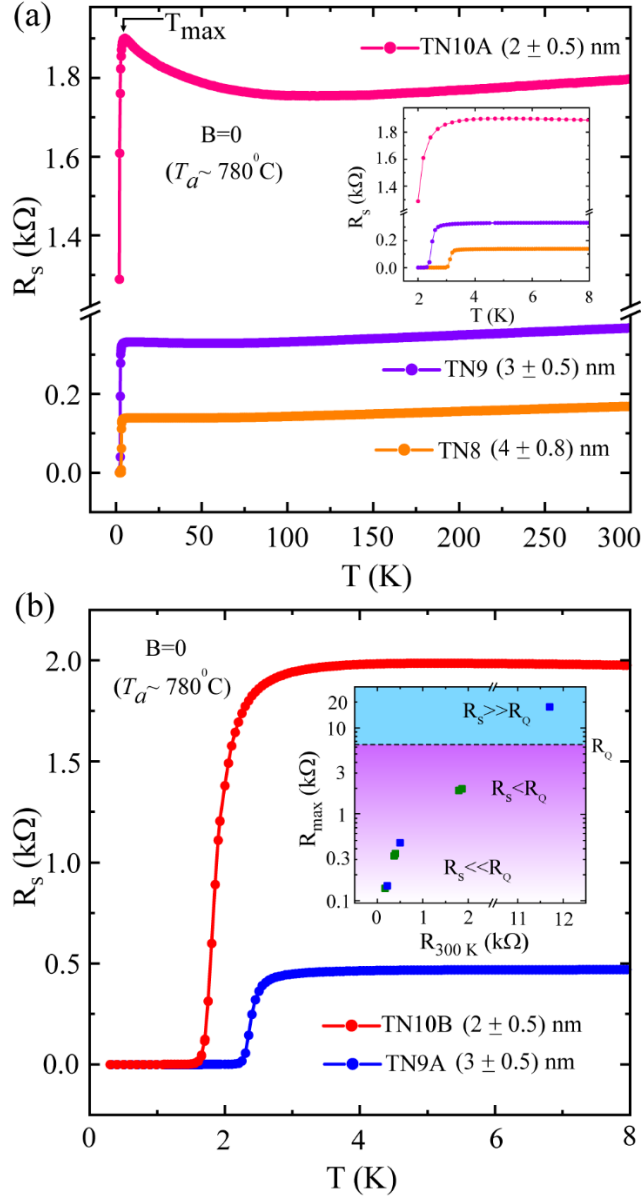

*Fig. 1*: *Temperature dependent sheet resistance $R_s(T)$ measured on TiN thin films with variation in film thickness. Based on the lowest achievable measurement temperature, two sets of samples are presented in (a) and in the main panel of (b). (a) $R_s(T)$ for the temperature range from room temperature down to 2K for three representative samples from the first set. Inset: the same set of $R_s(T)$ but for a selected range of temperature to highlight the metal-superconductor transition region. (b) $R_s(T)$ near the transition region for other two samples from the second set measured down to 300 mK. Here, the samples TN10A & TN10B were prepared in one single run and the samples TN9 & TN9A were grown in the same batch. Inset: The resistance $R_{max}$, measured at $T_{max}$ as shown by the arrow in Fig. 1(a) just above the superconducting onset, is plotted against the resistance measured at room temperature ($R_{300\,K}$) for all the samples. The dashed horizontal line, separating the blue and purple shaded regions, refers the quantum resistance $R_Q$. The resistance values for the samples are compared with respect to the quantum resistance.*

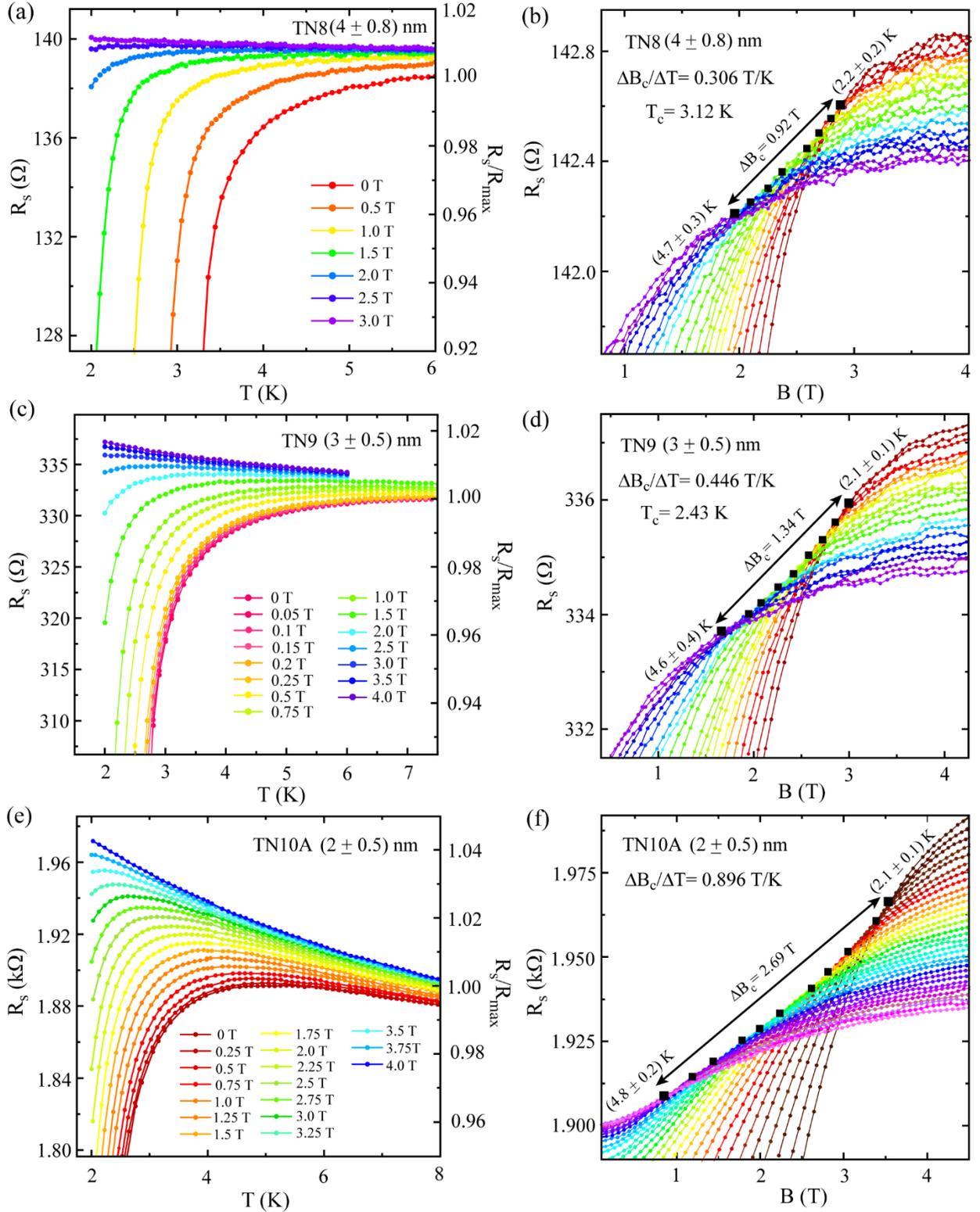

*Fig. 2:* Magnetic field dependent sheet resistance for the first set of samples (TN8, TN9A & TN10A) measured down to 2 K. Field dependent $R_s(T)$ for (a) TN8, (c) TN9A & (e) TN10A and isothermal magnetoresistance [$R_s(B)$] demonstrating multiple crossing points in the temperature range from 2 K to 5 K for (b) TN8, (d) TN9A & (f) TN10A. Here, the black solid diamonds mark the crossing points ($B_c$) for three consecutive $R_s(B)$ isotherms measured with an interval of 0.1 K in the temperature. Double sided arrows show the span $\Delta B_c$ in the field covering the multiple crossing region for the temperature window of 2-5 K. Here, it should be noted that a resistance change of about ~ 3 Ω between $R_s(T)$ and $R_s(B)$ for the sample TN8 occurs due to thermal cycling during the measurements.

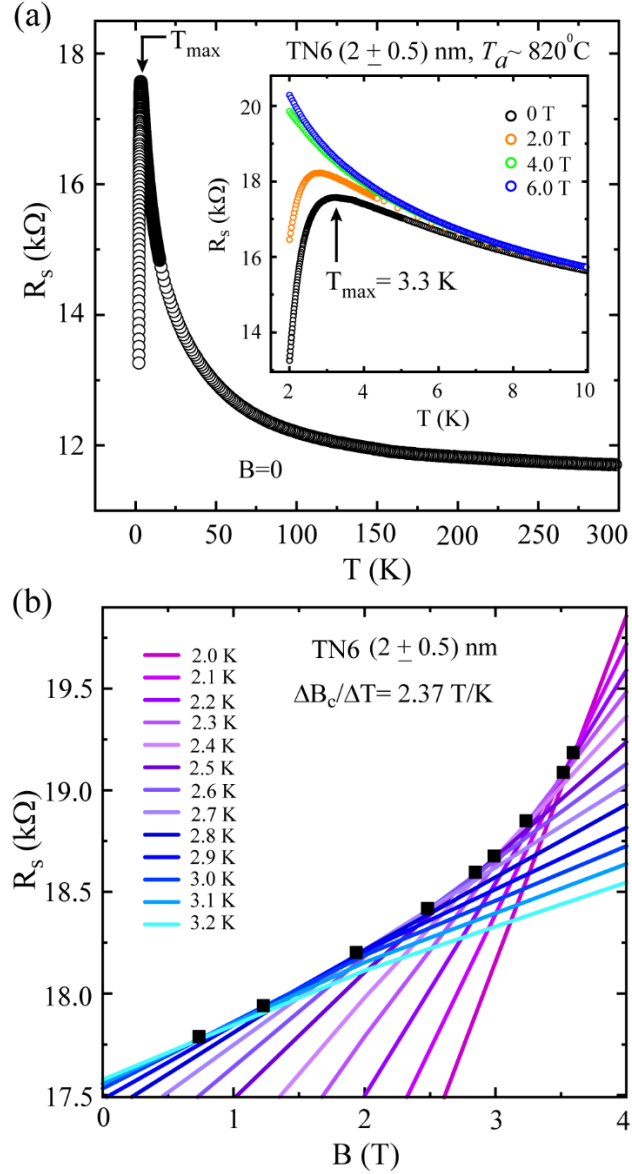

*Fig. 3: Field dependent $R_s(T)$ and $R_s(B)$ isotherms for a highly resistive TiN thin film sample (TN6) with $R_s >> R_Q$ in its normal state. The magnetoresistance $R_s(B)$ data is extracted from the field dependent $R_s(T)$ by the inverse matrix method. The thickness of the sample is $(2 \pm 0.5)$ nm and the annealing temperature $(T_a)$ during its growth was ~ 820°C. Here, the increased resistance is caused by the formation of elemental Si at $T_a$ ~820 °C in addition to TiN. (a) Zero-field $R_s(T)$ measured from room temperature down to 2 K. Inset: $R_s(T)$ measured under external magnetic field applied perpendicular to the sample plane. (b) Isothermal $R_s(B)$ obtained for the temperature window of 2 K to 3.2 K. The black filled diamonds represent the crossing points between the neighboring isotherms in small temperature intervals.*

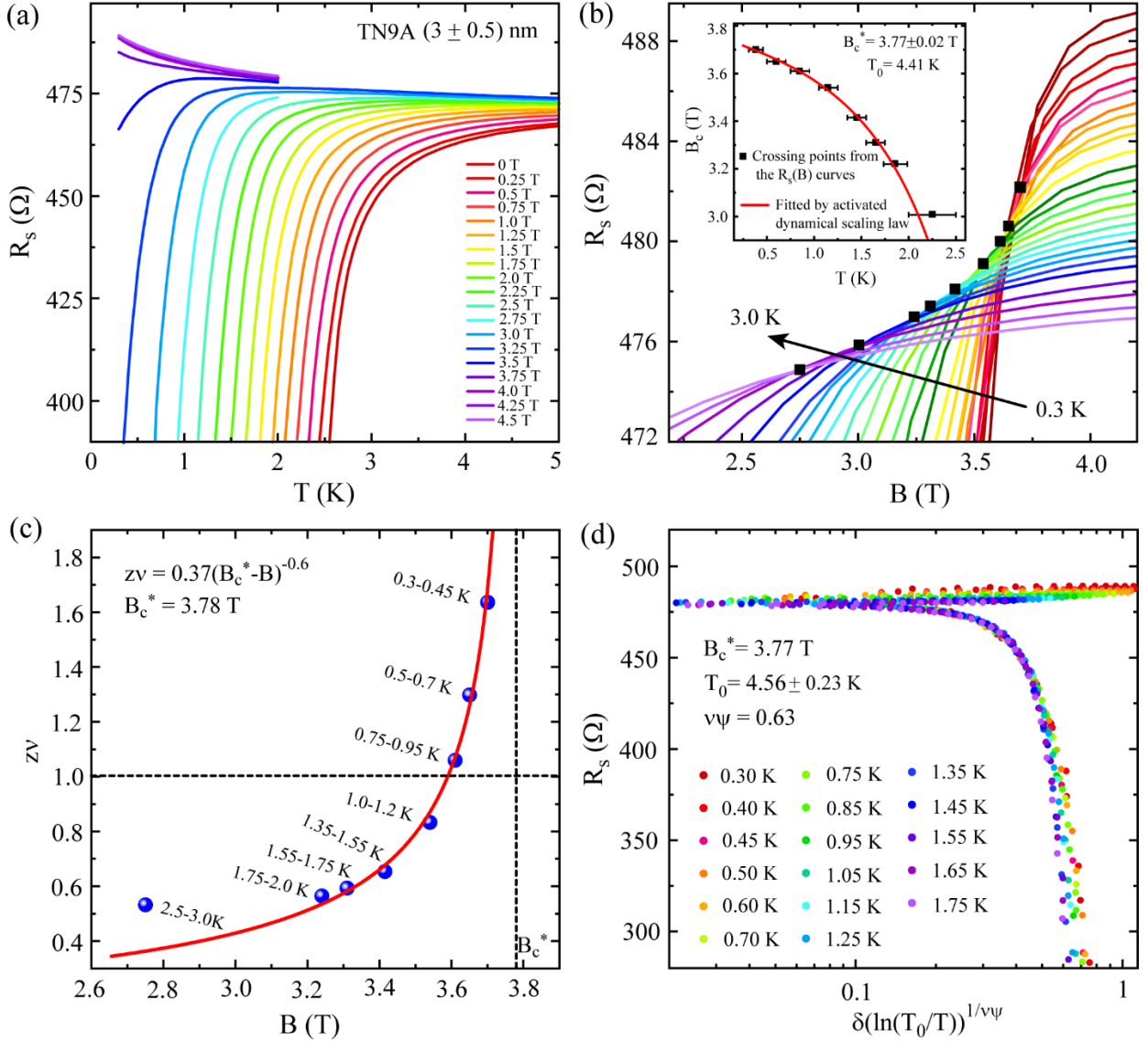

*Fig. 4*: *Probing quantum Griffiths singularity by magnetoresistance measurements carried out at lower temperature down to 300 mK for the sample TN9A (~3 nm). (a) A selected region of field dependent $R_s(T)$ measured under perpendicular magnetic fields from 0 to 4.5 T in 0.25 T steps. (b) Isothermal $R_s(B)$ measured in the temperature window of 0.3 K ≤ T ≤3 K. The magnetoresistance isotherms are measured with a temperature interval of 0.05 K between two neighboring isotherms from 0.3 K to 0.85 K, with 0.1 K interval from 0.85 K to 1.85 K and with 0.25 K interval from 2 K to 3 K. The crossing points are shown by the black diamonds. Inset: The crossing field ($B_c$) – temperature (T) phase boundary near the quantum superconductor-to metal transition. The red curve is the fit using Eq. (5) with $B_c^*$, u and p as adjustable parameters and the characteristic temperature $T_0$ fixed at 4.41 K (see the text for details). (c) Divergent behavior of critical exponent zν as a function of magnetic field. The zν values are obtained through finite size scaling (FSS) analysis for a set of adjacent magnetoresistance isotherms. The solid red curve is the fit based on activated scaling law using Eq. (2) and two black dashed lines represent zν = 1 (horizontal) & $B_c^*$ = 3.78 T (vertical). (d) Sheet resistance as a function of the scaling parameter $\delta(ln(T_0/T))^{1/\nu\psi}$ related to the activated dynamical scaling as described in Eq. (4) for the temperature span 0.3 K ≤ T ≤ 1.75 K. Error bars denote the temperature range for a set of magnetoresistance isotherms.*

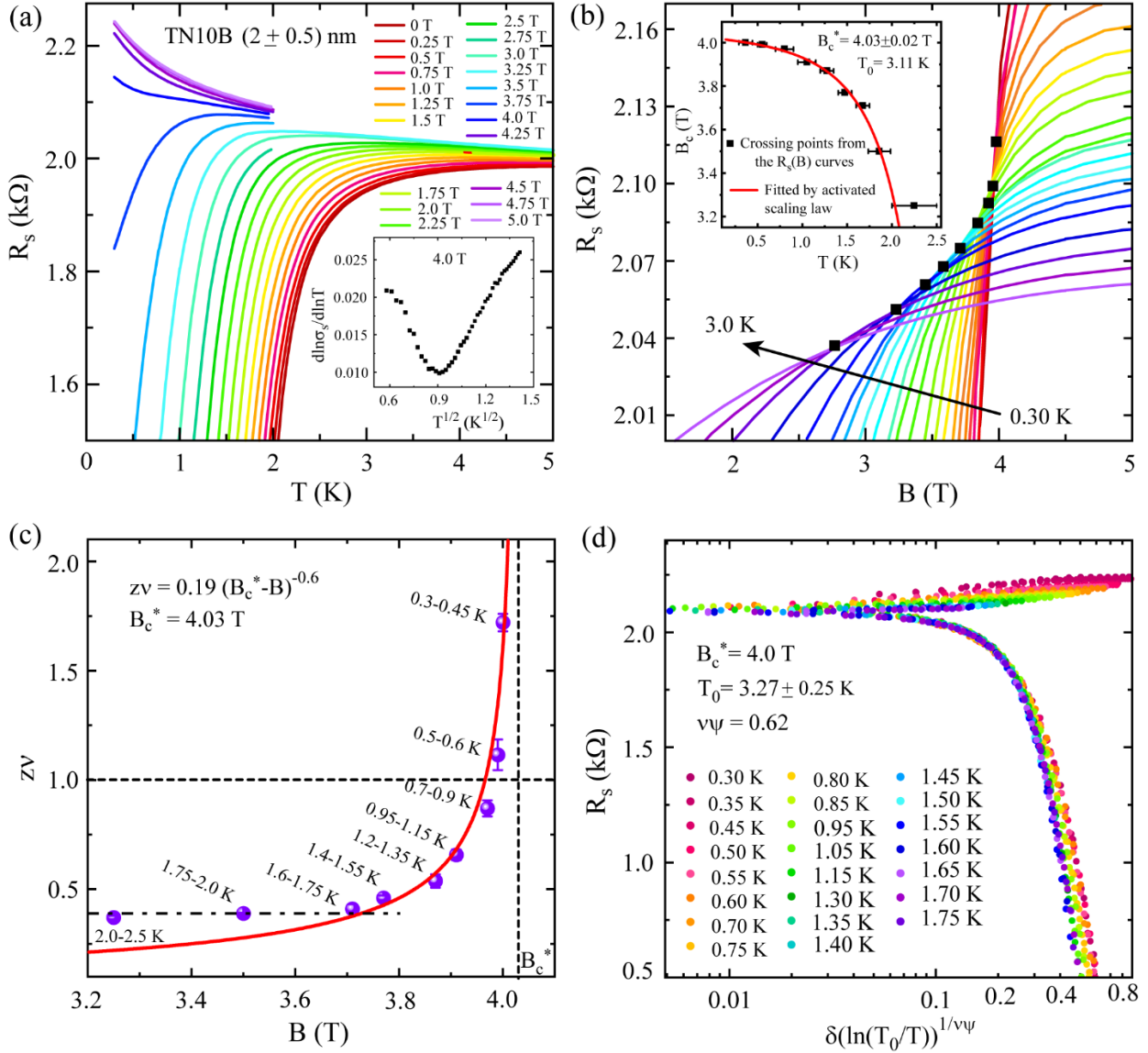

*Fig. 5*: Temperature and magnetic field dependent resistance $R_s(B,T)$ measurements carried out at low temperature down to 300 mK for the sample TN10B (thickness 2±0.5 nm). (a) A selected region of $R_s(T)$ measured under perpendicular magnetic fields from 0 to 5 T in 0.25 T steps. Inset: Logarithmic derivative of sheet conductance ($d\ln\sigma_s/d\ln T$) vs. $T^{1/2}$ obtained from the $R_s(T)$ measured under 4 T field. (b) A detailed view of isothermal $R_s(B)$ measured in the temperature window of 0.3 K ≤ T ≤3 K. The black diamonds represent the crossing points. Inset: The crossing field ($B_c$) – temperature (T) phase boundary near the quantum superconductor-to metal (SMT) transition. The red curve is the fit using Eq. (5) with $B_c^*$, $u$ and $p$ as adjustable parameters while the characteristic temperature $T_0$ was fixed at 3.11 K (see the text for details). (c) Field-dependence of the critical exponent product $zv$ which shows diverging behavior with magnetic field while approaching the characteristic critical field $B_c^*$ at the lowest measurement temperature. The $zv$ values are obtained through finite size scaling (FSS) analysis for a set of adjacent magnetoresistance isotherms in each small interval of temperature. The solid red curve is the fit based on the activated scaling law using Eq. (2). The black dashed lines represent $zv = 1$ (horizontal) & $B_c^* = 4.03$ T (vertical). The horizontal dashed-dotted broken line indicates a plateau region where $zv$ remains almost unchanged with changing magnetic field in the high temperature regime. (d) Sheet resistance as a function of the scaling parameter $\delta(\ln(T_0/T))^{1/\nu\psi}$ related to the activated dynamical scaling as described in Eq. (4) for the temperature span of 0.3 K ≤ T ≤ 1.75 K. Error bars denote the temperature range for a set of magnetoresistance isotherms.

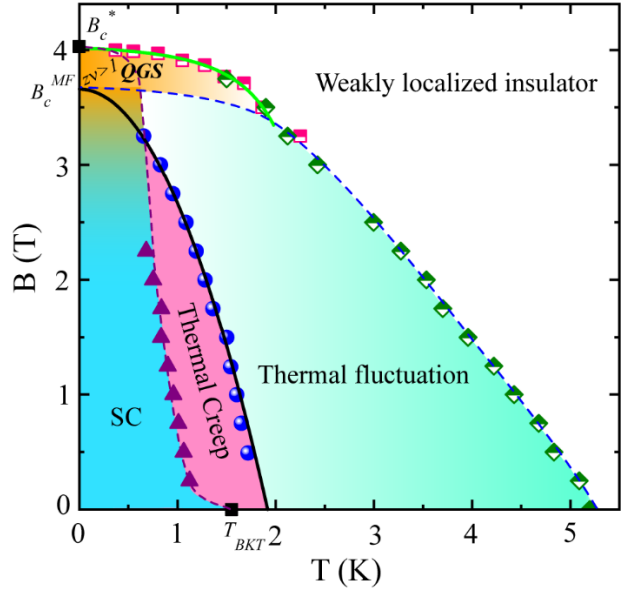

*Fig. 6:* *B-T Phase diagram constructed from the $R_s(B,T)$ measurements for the sample TN10B (~2 nm). The half-filled green diamonds represent the temperature $T_c^{Onset}$ corresponding to the temperature at which the slope $dR_s/dT$ changes its sign from negative to positive in the field dependent $R_s(T)$ while approaching the superconducting transition. The pink half-filled squares are the crossing critical fields $B_c(T)$ obtained from the crossing points of the neighboring MR isotherms in each small temperature interval. The blue spheres represent the mean-field critical field $B_c^{MF}(T)$ obtained from the field dependent $R_s(T)$ by using the Ullah-Dorsay scaling method for the field range 1.5 T ≤ B ≤ 3.25 T. For fields lower than 1.5 T, the $T_c$ values are considered at the place of $T_c^{MF}$. The purple triangles represent the characteristic temperature $T_{TAFF}$ which separates thermally activated flux flow (TAFF) regime from the quantum phase fluctuation regime (denoted here as SC regime) and is obtained from the Arrhenius plot analysis of the field dependent $R_s(T)$. The black solid curve is a fit for the $B_c^{MF}(T)$ using the empirical formula $B_c^{MF}(T) = B_c^{MF}(0)[1 - (T/T_c)^2]$. The green solid curve is the fit to the $B_c$-T phase boundary at the high field regime using activated dynamical scaling, as expressed in Eq. (5). Here the SC region, as separated from the thermal creep region by $T_{TAFF}$ (purple triangles), is extended up to the characteristic critical field $B_c^*$ in such a way that the extended region above the $B_c^{MF}$ corresponds to $zv \geq 1$. The blue broken curve following the $T_c^{Onset}$ (Green diamonds) is the guide to the eye which separates the weakly localized insulating state from the thermal fluctuation regime at high temperature (T > 1.75 K). At T<1.75 K, the green solid curve serves as the boundary between the weakly localized insulating region and the QGS region.*

*Supplementary Information*

# Emergence of quantum Griffiths singularity in disordered TiN thin films

Sachin Yadav,* [1,2] M.P. Saravanan, [3] and Sangeeta Sahoo* [1, 2]

[1]CSIR-National Physical Laboratory, Dr. K. S. Krishnan Marg, New Delhi-110012, India

[2]Academy of Scientific and Innovative Research (AcSIR), Ghaziabad- 201002, India

[3]Low Temperature Laboratory, UGC-DAE Consortium for Scientific Research, University Campus, Khandwa Road, Indore 452001, India

*Correspondences should be addressed to S. Y. (Email: sachin5459yadav@gmail.com) and

S. S. (Email: sahoos@nplindia.org)

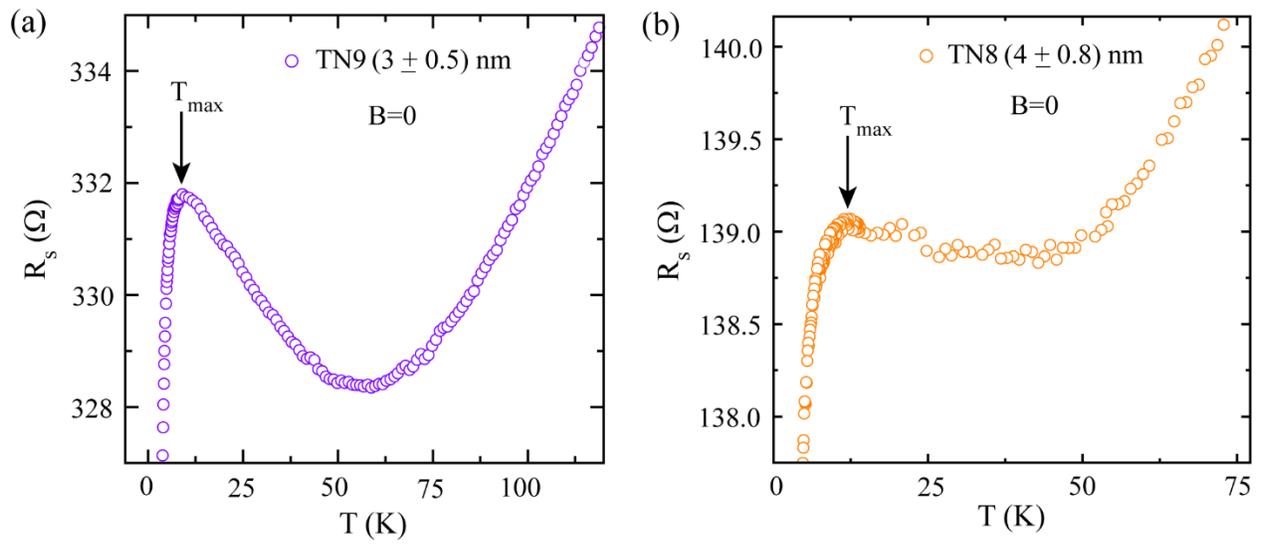

***Supplementary Fig. 1:*** *A set of zero field R(T) demonstrating a weakly localized metallic static before the onset of superconducting transition for the samples TN9 & TN8 annealed at 780 °C in (a) & (b), respectively. The resistance maximum/peak appearing at $T_{max}$ as marked by the arrows for both the samples.*

## Supplementary Note 1: Zero-field $R_s(T)$ for the samples TN5A (3 nm) & TN4 (4 nm) annealed at 820 °C

The zero field $R_s(T)$ measurements for the samples TN5A & TN4 in Supplementary Fig. 2 showcase a metal to superconductor transition (SMT) while lowering down the temperature down to 2K. The superconducting transition temperature ($T_c$), as defined by the temperature corresponding to the maximum $dR_s/dT$, appears as ~ 2.95 K & ~ 3.60 K for the samples TN5A & TN4, respectively.

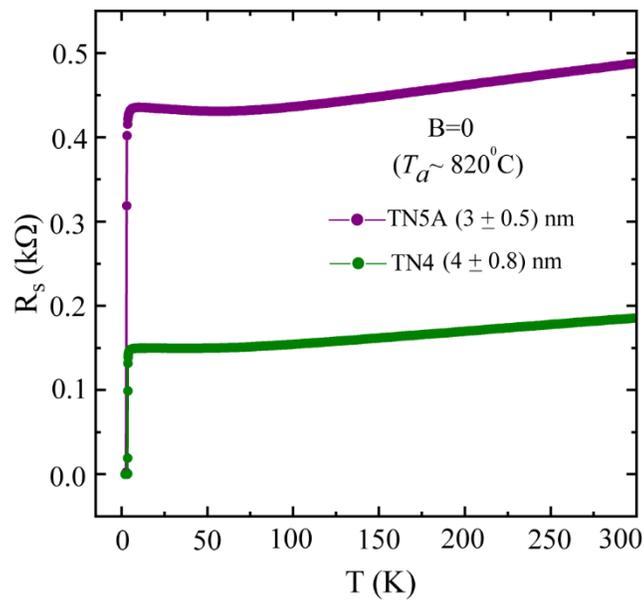

*Supplementary Fig. 2: Zero field temperature dependent sheet resistance $R_s(T)$ measurements from room temperature (300 K) down to 2 K for the samples TN5A (thickness ~3 nm) & TN4 (thickness ~4 nm) that were annealed at 820 °C.*

## Supplementary Note 2: Field dependent $R_s(T)$ curves and magnetoresistance isotherms for the samples TN5A & TN4

The $R_s(T)$ curves for the sample TN4 show metal to superconductor phase transition, while lowering the temperature down to 2 K in the absence of magnetic field. Under the application of perpendicular magnetic field, superconducting transition shifts monotonically toward the lower temperature side as shown in Supplementary Fig. 3(a). Further, increment in the magnetic field transforms the superconductor into metal, indicating towards the field induced superconductor to metal transition (SMT). The behavior of the SMT for the sample TN4 is further explored by the measurement of magnetoresistance isotherms shown in Supplementary Fig. 3(b). The magnetoresistance isotherms demonstrate multiple crossing points that are different from the conventional systems exhibiting superconductor to insulator transition (SIT) with a single crossing point. Here, black solid diamonds mark the multiple crossing points, where each crossing point acts as a single critical point for a set of neighboring magnetoresistance isotherms in small temperature interval as shown in Supplementary Fig. 3(b). Further, we have observed similar features on the sample TN5A with reduced film thickness as shown in Supplementary Fig. 3(c-d). Initially, with the increase in the magnetic field the $R_s(T)$ curves of the sample TN5A shift monotonically towards the lower temperature regime and gradually transform into weakly localized metallic state with further increment in the magnetic field and a field induced SMT occurs as shown in Supplementary Fig. 3(c). Further, this SMT behavior is explored by isothermal magnetoresistance measurements as shown in Supplementary Fig. 3(d), where the $R_s(B)$ curves demonstrate multiple crossing points (MCPs) with the variation in temperature and the MCPs are marked by the black solid diamonds. These MCPs form a continuous curve, which is a signature of quantum Griffiths singularity (QGS). Therefore, magnetoresistance isotherms of both the samples (TN4 & TN5A) in Supplementary Fig. 3(b-d) demonstrate the appearance of MCPs that form a continuous curve indicating towards the presence of QGS.

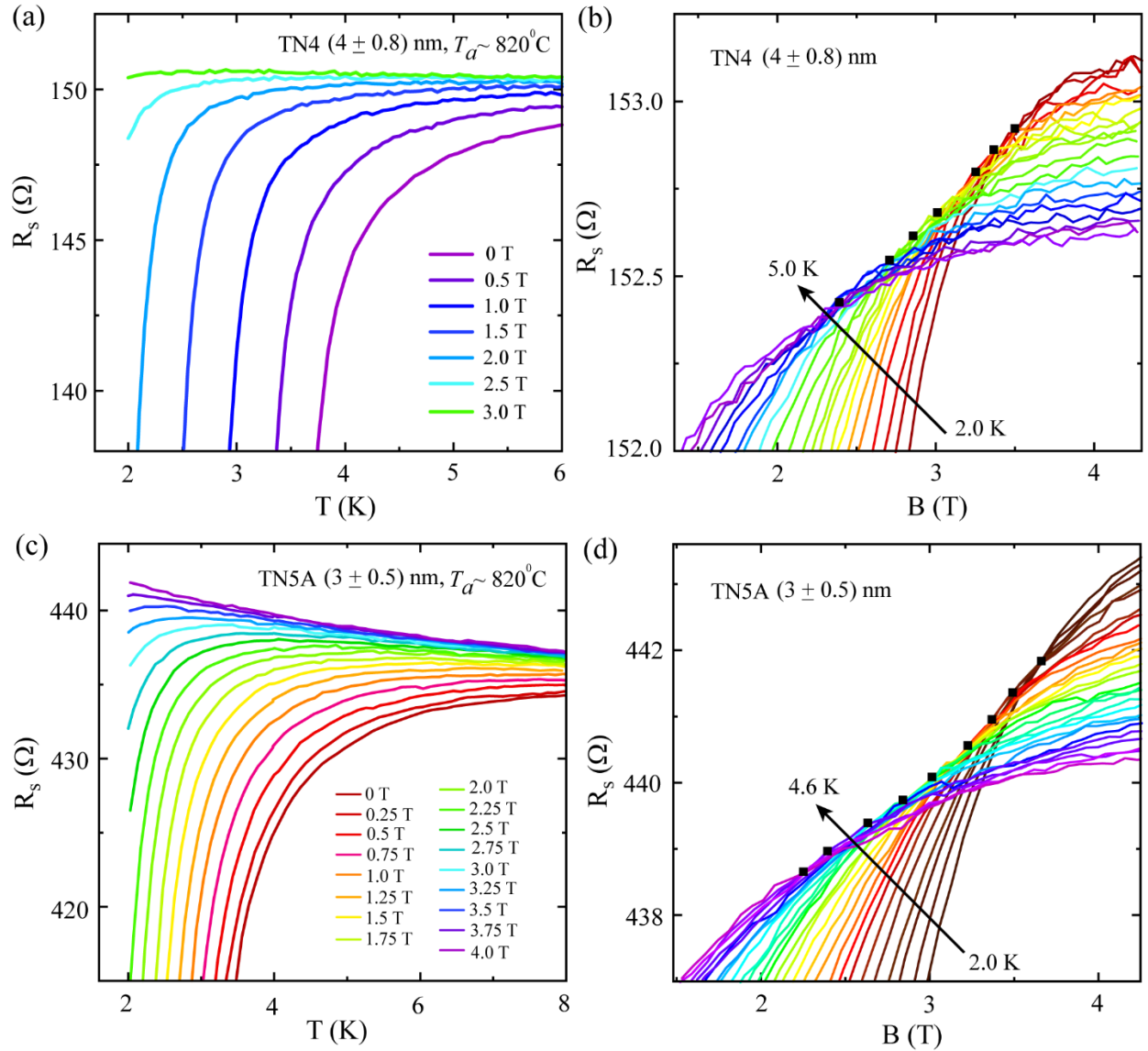

***Supplementary Fig. 3:*** *The temperature and field dependent sheet resistance $R_s(B,T)$ measured under perpendicular magnetic field for the samples TN4 & TN5A. An enlarged view of $R_s(T)$ measured under different magnetic fields for (a) TN4 (~4 nm) and (c) TN5A (~3 nm). A detailed view of the magnetoresistance $R_s(B)$ isotherms near the field induced transition for (b) TN4 and (d) TN5A. The temperature windows for the measurement of $R_s(B)$ isotherms are 2 - 5 K and 2-4.6 K for TN4 and TN5A, respectively. The solid black arrows indicate the temperature variation from 2 K to 5.0 K & 4.6 K for the $R_s(B)$ isotherms for samples TN4 & TN5A, respectively. The solid black diamonds mark the crossing points between the adjacent magnetoresistance isotherms in each small interval of temperature for both the samples. Here, it should be noted that a resistance change of about ~ 3 Ω between $R_s(T)$ and $R_s(B)$ for the sample TN5A occurs due to thermal cycling during the measurements.*

## Supplementary Note 3: Finite size scaling (FSS) analysis for TN9A

*Finite size scaling analysis:*

To investigate the existence of quantum Griffiths singularity (QGS) in TiN thin films, we have performed the finite size scaling (FSS) analysis on the magnetoresistance isotherms in the vicinity of QCP which can be expressed as [1],

$$R_s(B,T) = R_c f[(B - B_c)t] \qquad \text{(S1)}$$

$$t = (T/T_0')^{-1/zv} \qquad \text{(S2)}$$

where, $R_c$ & $B_c$ correspond to the critical resistance & critical magnetic field values, respectively, at the crossing point for a given set of $R_s(B)$ curves. $f(x)$ is the scaling factor with $f(0) = 1$, $z$ and $v$ are the dynamical and static critical exponents, respectively. $T_0'$ is the lowest temperature for each set of magnetoresistance isotherms associated with a particular crossing point. During the scaling procedure, the critical exponent product $zv$ was obtained by assuming that the functional form of scaling factor $f(x)$ remains unchanged for $x > 0$ and $x < 0$. Firstly, we have formulated different sets of neighboring $R_s(B)$ isotherms in small interval of temperature where the isotherms from each set cross at a single point which corresponds to the critical magnetic field is $B_c$ and the critical resistance $R_c$. The normalized resistance ($R_s/R_c$) when plotted against the scaling variable $(B - B_c)t$, the magnetoresistance isotherms from a particular set collapse on top of each other by adjusting the scaling parameter ($t$). The obtained value of $t$ from the scaling collapse is plotted as $\ln(t)$ versus $\ln(T/T_0')$ where $T_0'$ is the lowest temperature in the particular set and the scaling parameter $t = 1$ for the isotherm corresponding to the temperature $T_0'$. A linear fit yields the slope $= -1/zv$ from where the value of effective critical exponent $zv$ can be obtained.

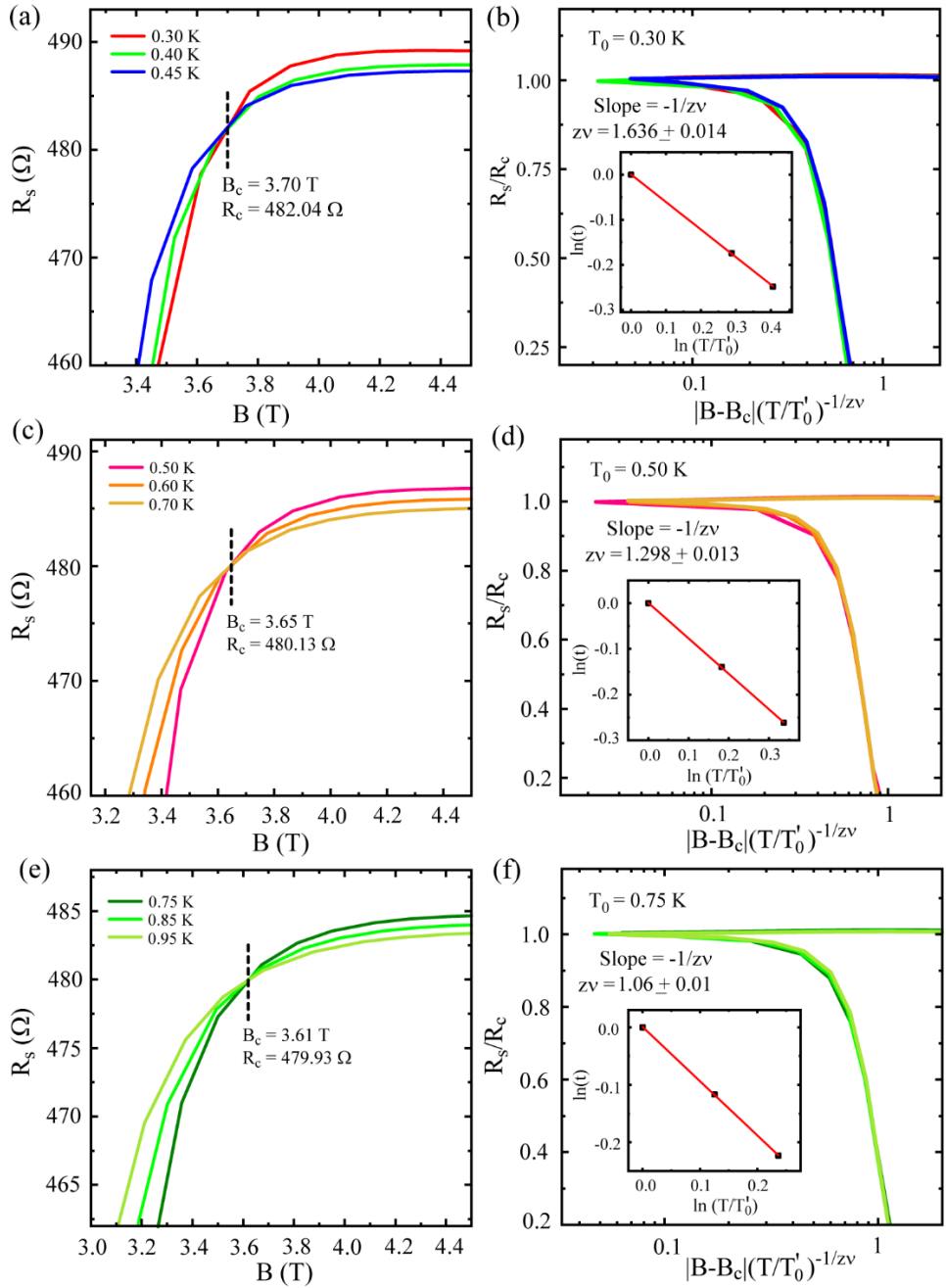

***Supplementary Fig. 4:*** *Finite size scaling (FSS) analysis for the sample TN9A (presented in the main manuscript) for the temperature window of 0.3 K- 0.95 K. (a), (c) & (e) The variation of sheet resistance ($R_s$) as a function of magnetic field for three sets of temperature points (0.3 K- 0.95 K) near the transition regime. (b), (d) & (f) The normalized sheet resistance as a function of the scaling variable $|B - B_c|(T/T_0')^{-1/z\nu}$. Insets: linear fits to ln(t) versus ln(T/T_0') for obtaining the slopes. Here, t is the scaling factor and $T_0'$ is the lowest temperature in each of the sets of $R_s(B)$ curves for a particular crossing point. The inverse of each slope gives the value of the critical exponent product $z\nu$ for the set of temperature points related to a particular crossing point. Here, $B_c$ corresponds to a a critical field as determined by the crossing point for the set of magnetoresistance isotherms and $T_0$ is the lowest temperature in each of the set of $R_s(B)$ curves for a particular crossing point.*

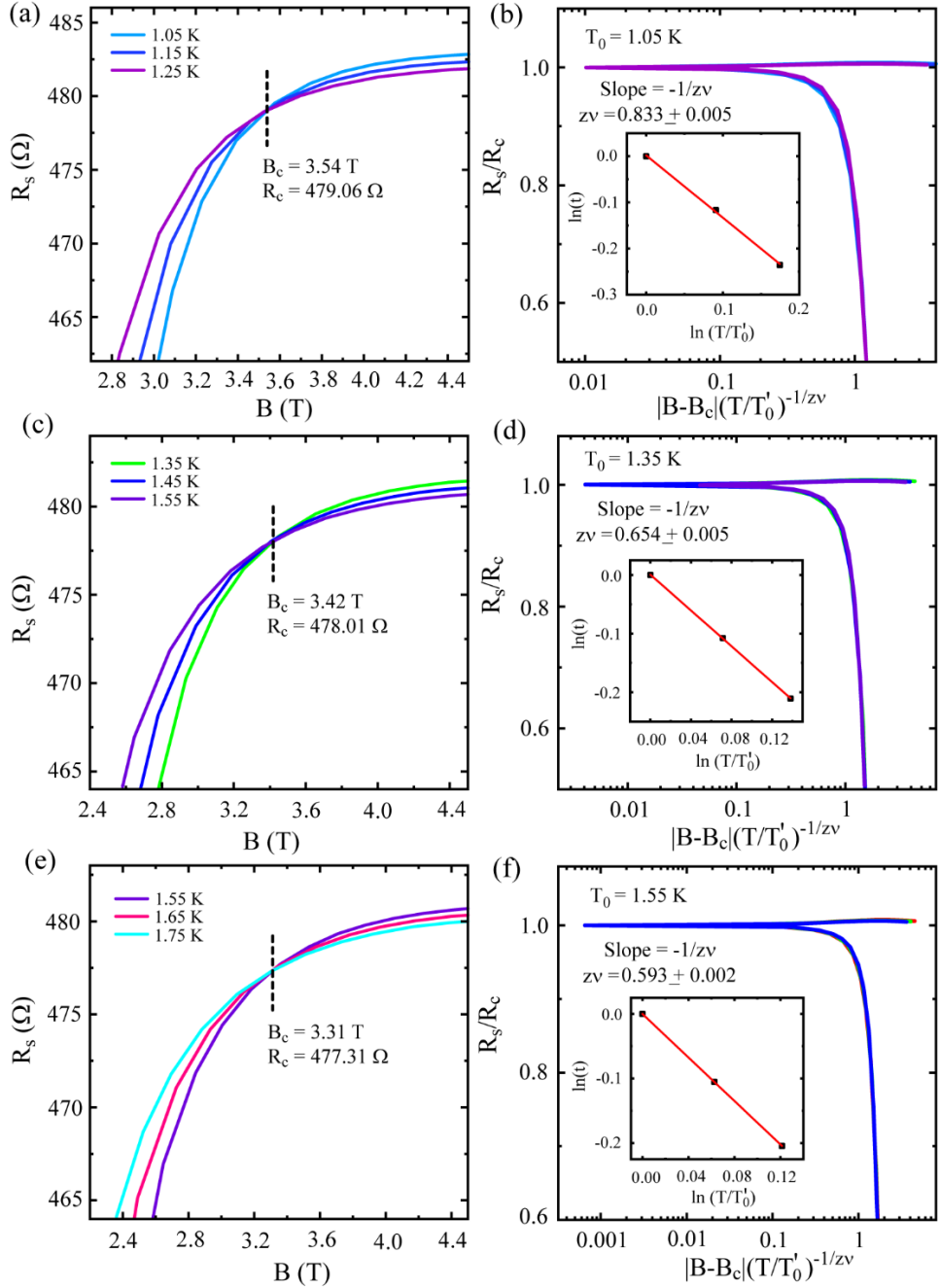

***Supplementary Fig. 5:*** *Finite size scaling (FSS) analysis for the sample TN9A for the temperature range 1.05 K -1.75 K. (a), (c) & (e) The variation of sheet resistance ($R_s$) as a function of magnetic field for three sets of temperature points (1.05 K- 1.75 K) near the transition regime. (b), (d) & (f) The normalized sheet resistance as a function of the scaling variable $|B - B_c|(T/T_0')^{-1/z\nu}$. Insets: linear fits to ln(t) versus ln(T/T_0') for obtaining the slopes. Here, t is the scaling factor and $T_0'$ is the lowest temperature in each of the sets of $R_s(B)$ curves for a particular crossing point. The inverse of each slope gives the value of the critical exponent product zν for the set of temperature points related to a particular crossing point. Here, $B_c$ corresponds to a critical field as determined by the crossing point for the set of magnetoresistance isotherms.*

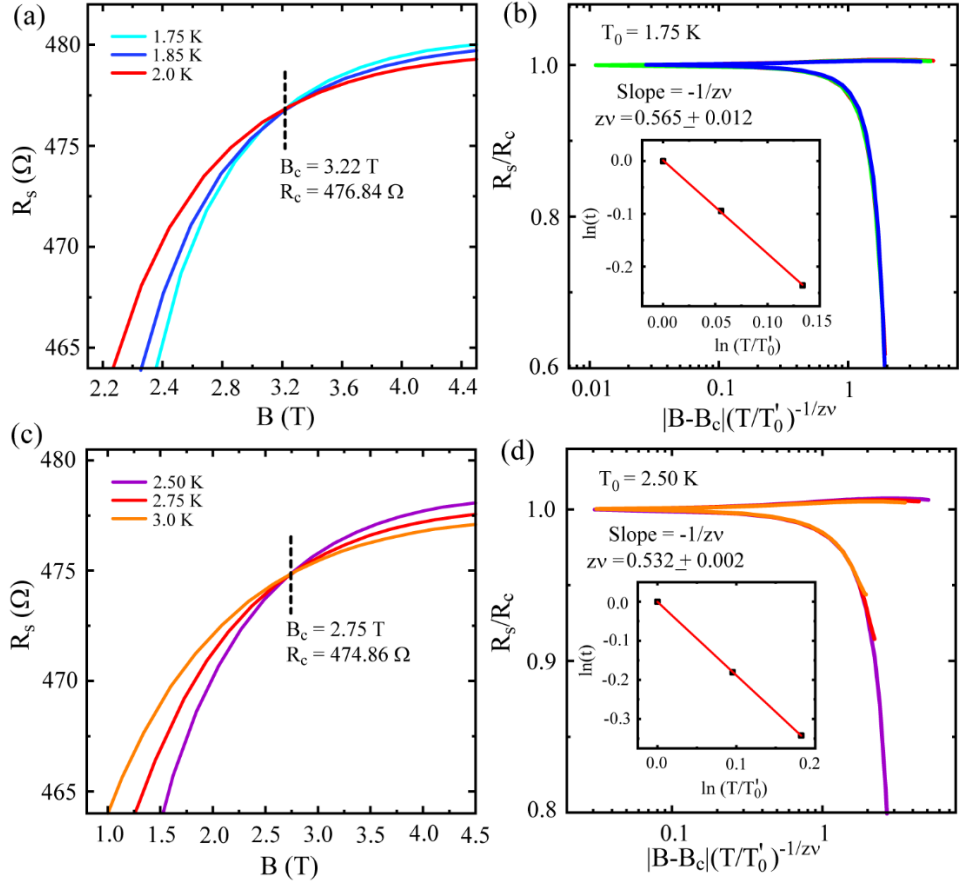

***Supplementary Fig. 6:*** *Finite size scaling (FSS) analysis for the sample TN9A for the temperature range 1.75 K - 3.0 K. (a), (c) & (e) The variation of sheet resistance ($R_s$) as a function of magnetic field for three sets of temperature points (1.75 K- 3.0 K) near the transition regime. (b), (d) & (f) The normalized sheet resistance as a function of the scaling variable $|B - B_c|(T/T_0')^{-1/z\nu}$. Insets: linear fits to ln(t) versus ln(T/$T_0'$) for obtaining the slopes. Here, t is the scaling factor and $T_0'$ is the lowest temperature in each of the sets of $R_s(B)$ curves for a particular crossing point. The inverse of each slope gives the value of the critical exponent product zν for the set of temperature points related to a particular crossing point. Here, $B_c$ corresponds to a critical field as determined by the crossing point for the set of magnetoresistance isotherms.*

## Supplementary Note 4: Variation of critical exponent *zν* as a function of temperature for the samples TN9A & TN10B

The dynamical critical exponent *zν* obtained from the FSS analysis is plotted as a function of temperature for the samples TN9A & TN10B and the plots are shown in Supplementary Fig. 7 (a) & (b) respectively. Here, temperature corresponding to a particular *zν* is obtained by averaging the temperatures of adjacent isotherms that generate the crossing point for a given set. In case of activated dynamical scaling, the effective value of critical exponent *zν* can be expressed by the infinite-randomness critical exponents υψ and at $T \rightarrow 0$, the two types of critical exponents, namely, *zν* and υψ are related as [2],

$$\left(\frac{1}{\nu z}\right)_{eff} = \frac{1}{\upsilon \psi} \frac{1}{\ln(T_0/T)} \qquad (S3)$$

As, $T \rightarrow 0$, the right-hand side of the above-mentioned expression vanishes and *zν* is expected to diverge in the zero temperature limit. The blue solid curves in Supplementary Fig. 7 are the fits to the experimental data by using the expression given in Eq. S3. The fits follow the experimental data nicely up to a temperature as high as ~ 1.75 K for both the samples. However, a deviation at the lowest measurement temperature is observed particularly for the more disordered sample TN10B. The reason of the deviation is not clear at present. A more detailed study at much lower temperature with smaller interval in temperature is required for a clearer understanding of the behavior of *zν* at the zero-temperature limit.

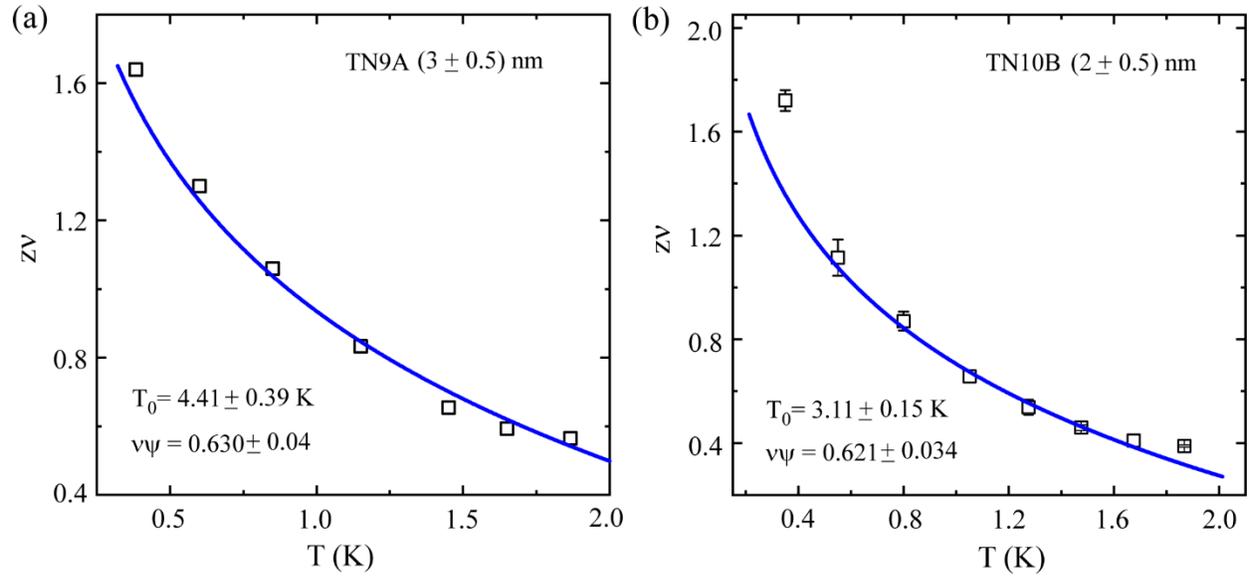

***Supplementary Fig. 7:*** *The variation of dynamical critical exponent zv as a function of temperature for the samples (a) TN9A & (b) TN10B . Blue solid curves are the fits to the data by using Eq. (S3) with $v\psi$ & $T_0$ are adjustable parameters. The fitted curves have yielded the values of $v\psi$ & $T_0$ for the samples TN9A & TN10B and the corresponding values for TN9A & TN10B are mentioned in (a) & (b) respectively.*

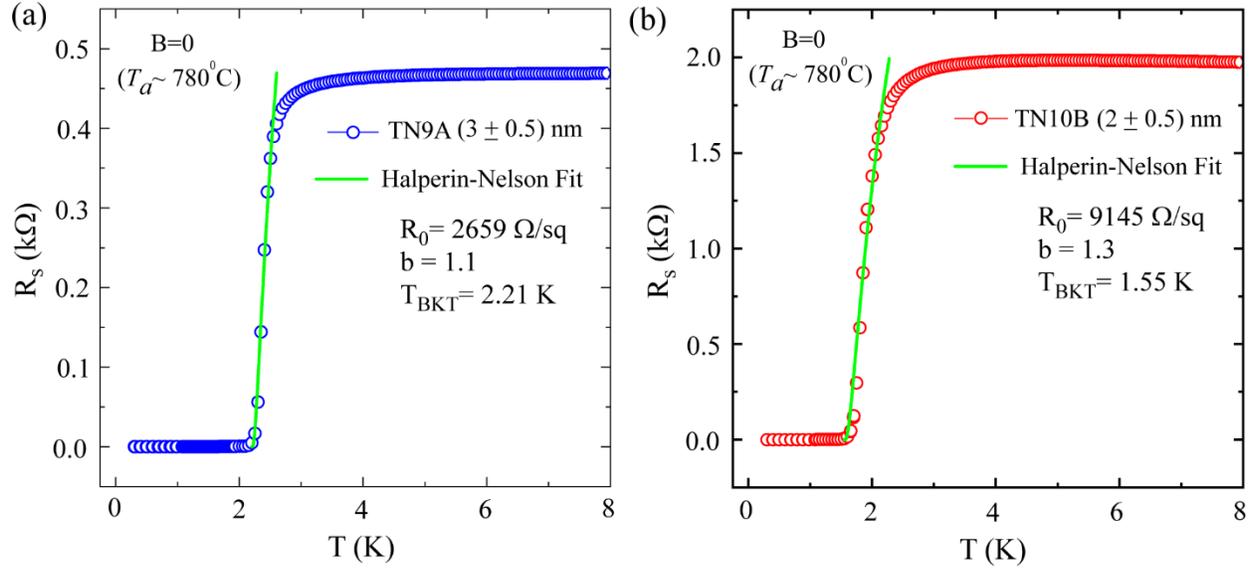

***Supplementary Fig. 8:*** *Extraction of $T_{BKT}$ for the samples TN9A & TN10B by using the Halperin-Nelson model. The sheet resistance as a function of temperature in the absence of magnetic field for the samples TN9A in (a) & TN10B in (b). The open circles are the experimental data, whereas solid green curves are the least - square fits obtained from the Halparin-Nelson formula of*

$R_s = R_0 exp[-b(T/(T/T_{BKT} - 1)^{-1/2}]$ . *The obtained values of $R_0$, b and $T_{BKT}$ are mentioned in (a) & (b) for the sample TN9A & TN10B, respectively.*

**Supplementary Note 5: Estimation of mean field upper critical field $B_c^{MF}(T)$ by using Ullah- Dorsey (UD) scaling of fluctuation conductivity for the samples TN9A & TN10B**

One of crucial components for establishing a comprehensive phase diagram for a 2D superconductor is the mean field upper critical field $B_{c2}^{MF}(T)$ which separates the thermally driven amplitude fluctuation regime from the thermally activated phase fluctuation by means of vortex motion regime. Here, we adopted the Ullah-Dorsey scaling theory [3-5] to estimate the mean field upper critical field $B_{c2}^{MF}(T)$ from the fluctuation conductivity $G_{fl}^{2D}$ which can be calculated from the field dependent $R_s(T)$ as, $G_{fl}^{2D} \equiv 1/R_s(T) - 1/R_N(T)$, where $R_N(T)$ is the temperature dependent normal state resistance. The fluctuation conductivity $G_{fl}^{2D}$ is the excess conductivity due to the thermal superconducting fluctuations. In a finite magnetic field $B$, $G_{fl}^{2D}$ can be expressed under the Hartree approximation as [6],

$$G_{fl}^{2D}\left(\frac{B}{T}\right)^{1/2} = F\left(\frac{T-T_c^{MF}(B)}{(TB)^{1/2}}\right), \quad F(x) \propto \begin{cases} -x \ (x \ll 0) \\ x^{-s} \ (x \gg 0) \end{cases} \quad (S4)$$

where, $F(x)$ is the scaling parameter, $T_c^{MF}(B)$ is the mean field critical temperature in a field, and the exponent $s = 1$ in 2D superconductors. The $T_c^{MF}(B)$ is used as an adjusting parameter. $R_N(T)$ is obtained by the extrapolation of a linear fit to the $R_s(T)$ considered in the tempearture range far above the transition temperature so that superconducting fluctuations can be ignored in the $R_N(T)$. By using the UD scaling theory, as expressed in Eq. S4, $T_c^{MF}(B)$ is obtained by adjusting it to achieve for the best collapse for $T > T_c^{MF}(B)$ and with a slope of -1, when $G_{fl}^{2D}(T)$ is plotted in a log-log scale as shown in Supplementary Fig. 9. We have plotted $G_{fl}^{2D}\left(\frac{B}{T}\right)^{1/2}$ versus $\frac{T-T_c^{MF}(B)}{(TB)^{1/2}}$ in log-log scale for the samples TN9A and TN10B in Supplementary Fig. 9(a) & (b), respectively. At temperature above the $T_c^{MF}(B)$ (the lower branch) and for the field range 1.0 to 2.75 T, the $G_{fl}^{2D}(T)$ curves collapse onto a single one with a slope of -1 for the sample TN9A and the field range for the sample TN10B is 1.5 to 3.25 T. Finally, the $B_{c2}^{MF}(T)$ curve is derived by using $T_c^{MF}(B)$ for a particular field as obtained by the UD scaling explained in Eq. S4. We found that the obtained $T_c^{MF}(B)$ values closely match with the $T_c$ values as defined by the criterion of $dR_s/dT$ maximum.

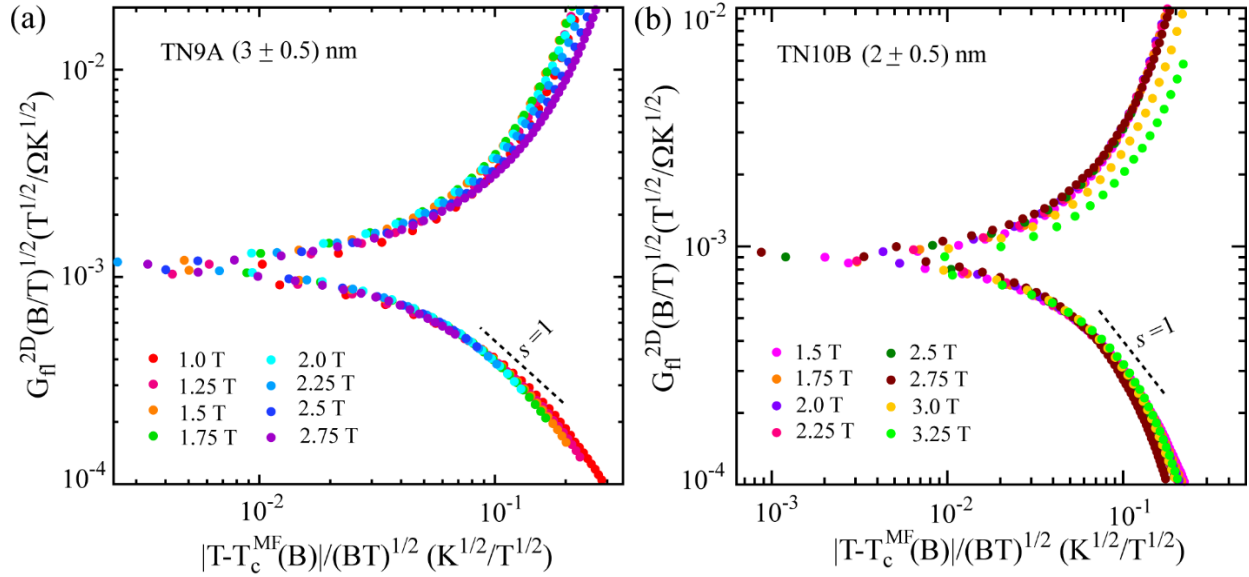

**Supplementary Fig. 9:** *Ullah-Dorsey (UD) scaling of the fluctuation conductivity $G_{fl}^{2D}$ for the estimation of the mean field upper critical field $B_{c2}^{MF}(T)$ for samples TN9A in (a) and TN10B in (b). The fluctuation conductivity $G_{fl}^{2D}$ is obtained from the field dependent $R_s(T)$. The best collapse corresponding to the exponent $s = 1$ from the Eq. (4) occurs for the field range 1.0 to 2.75 T and 1.5 to 3.25 T for TN9A and TN10B, respectively.*

**Supplementary Note 6: Evaluation of the nature of the insulating phase for the samples (TN10A, TN10B and TN6) showing field-induced SIT behavior**

The nature of the insulating phase for the samples showing magnetic-field-induced SIT at the lowest temperature has been investigated by magnetoresistance (MR) measurements at higher field which is shown in Supplementary Fig. 10. The MR isotherms presented in Supplementary Fig. 10(a) for the sample TN10A demonstrate a non-monotonic behaviour that consists of a positive MR (at low fields) & negative MR (at high fields). Similar behaviour has been observed in the literature for TiN thin films, where the positive MR is due to the localized Cooper pairs & negative MR emerges from the enhanced conductance due to the breaking of Cooper pairs under high magnetic field [7-10]. Further, it has been predicted [11] and also experimentally observed [12,13] that the critical resistance ($R_c$) at the quantum critical point (QCP) in the field tuned SIT is equivalent to the quantum resistance of Cooper pairs $R_Q=h/(2e)^2=$ 6.45 k$\Omega$ [11]. For a detailed understanding of the nature of the insulating phase near the quantum transition, in Supplementary Fig. 10(b), we have presented the $R_s(B)$ measurements carried out on the sample TN10B at much lower temperature down to 300 mK and up to a field of about 10 T. The sample TN10B in Supplementary Fig. 10(b) also demonstrates the similar non-monotonic MR isotherms where in the close vicinity of the critical field for $B > B_c^*$, resistance increases sharply and reaches to a maximum at $B_{max}$. With further increasing the field for $B>B_{max}$, resistance decreases slowly and the downward slope leads to negative magnetoresistance. Here, the appearance of MR peak above the critical field for the field-induced SIT indicates the presence of localized Cooper pairs [7,8,10] and the peak formation occurs due to the competition between fermionic and bosonic contributions. The strong positive MR region for $B_c^*<B<B_{max}$ indicates the field induced suppression of the long-range phase coherence and the localization of Cooper pairs [7,10]. While on the higher field side, the negative MR indicates the enhanced conductance in the quasiparticle channel due to the pair breaking. Further, the difference in the critical resistance $R_c$ measured at $B_c^*$ and the $R_N$ is very small (~161 $\Omega$) and the critical resistance $R_c$ (~2146 $\Omega$ ) is very close to the $R_N$ but far from the quantum resistance of Cooper pairs. Therefore, the fermionic contributions can be significant in the insulating state near the transition [14,15]. For a higher resistive sample TN6 presented in

Supplementary Fig. 10(e), we observe similar type of broad peak in the MR with the critical resistance being close to the normal state resistance but far from the quantum resistance of Cooper pairs. Hence, these observations indicate the dominant presence of the fermionic contribution in addition to the localized Cooper pairs in the insulating state near the transition for $B>B_c^*$.

Further, we have analyzed the isomagnetic $R_s(T)$ data for various magnetic fields in the surrounding of the critical field. As in 2D, the logarithmic temperature dependence of conductance mainly indicates the weak localization (WL) effects and electron-electron interaction (EEI) in the fermionic channel [14,16], here, we have presented the logarithmic temperature dependence of the conductance $G(T) =1/R(T)$ for the sample TN10B in Supplementary Fig. 10(c-d) and for TN6 in Supplementary Fig. 10(f). The magnetic field positions are marked in the respective $R(B)$ curves presented in Supplementary Fig. 10(b) & (e) for the samples TN10B and TN6, respectively. For the sample TN10B, the conductance versus log($T$) dependences for the regions of positive MR and the high field negative MR are separately presented in Supplementary Fig. 10(c) & (d), respectively. Below the critical field with lowering temperature, conductance deviates from the log($T$) dependence at temperature [shown by the black arrows in Supplementary Fig. 10(c)] where conductance starts to rise due to the formation of Cooper pairs and the conduction is dominated by the superconducting fluctuations in the Cooper channel. For the field just above the critical field up to the $B_{max}$, the conductance follows the log($T$) dependence for much wider temperature range and the deviation is observed at low temperature. The deviation from the log($T$) dependence at the lower temperature in the positive MR region indicates the presence of localized Cooper pairs. However, at $B= B_{max}$, the logarithmic temperature dependence of the conductance is observed almost all the way down to the lowest temperature. Here, the close association of the high-temperature behaviors among the isomagnetic conductance curves presented in Supplementary Fig. 10(c) & (f) indicates the same origin of carriers in the single particle and Cooper channel. Further for $B>B_{max}$, a complete logarithmic temperature dependence of conductance is observed which indicates the fermionic nature of the conduction at higher field side with negative MR region as shown in Supplementary Fig. 10(d) for the sample TN10B.

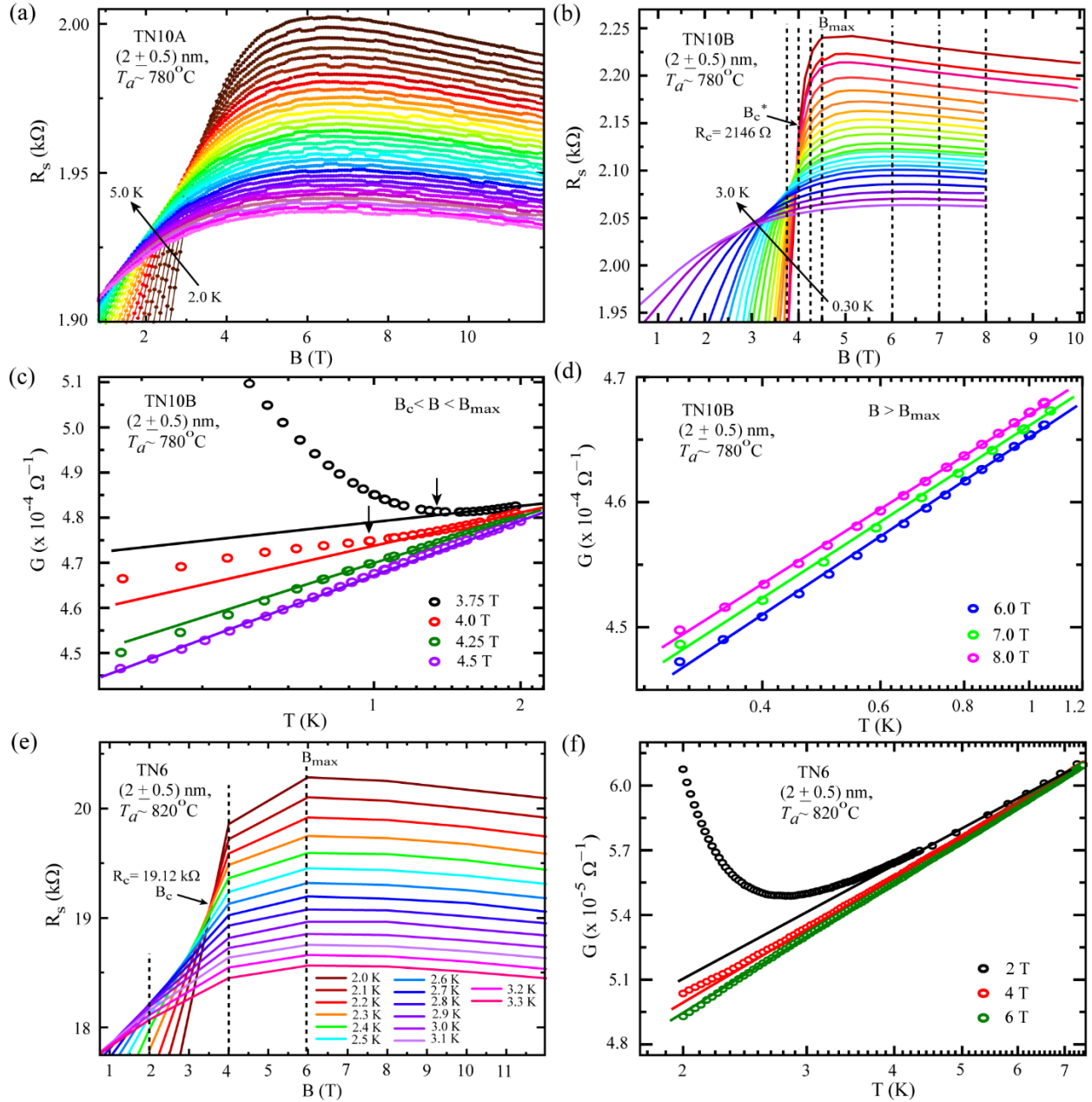

***Supplementary Fig. 10:*** (a) & (b) The field dependent resistance $R_s(B)$ isotherms demonstrating the appearance of MR peaks that show a crossover from positive to negative MR at high magnetic field for the samples TN10A and TN10B, respectively. The temperature intervals for the $R_s(B)$ isotherms are 2-5 K for TN10A and 3-0.3 K for TN10B. The dotted vertical lines in (b) for TN10B represent specific field values at which temperature dependence of conductance ($G=1/R_s$) data are analyzed further. (c) & (d) The temperature dependence of conductance in logT scale measured at the fields that are marked in (b) for $B \leq B_{max}$ & for $B>B_{max}$, respectively. At higher field side with the negative MR region, the logarithmic temperature dependence indicates the fermionic nature of the electronic interaction and localization. (e) & (f) A similar type of analysis for the sample TN6 by using $R_s(B)$ isotherms and temperature dependence of conductance for specific fields, respectively. Here, the $R_s(B)$ isotherms for TN6 also demonstrate a broad resistance peak at a crossover field which separates the negative MR region at high magnetic field from the positive MR region near the transition. The logT dependence for the conductance is also evident in (f) for high field.

**Supplementary References:**

1. Sondhi, S. L., Girvin, S. M., Carini, J. P. & Shahar, D. Continuous quantum phase transitions. *Reviews of Modern Physics* **69**, 315-333 (1997).

2. Lewellyn, N. A. *et al.* Infinite-randomness fixed point of the quantum superconductor-metal transitions in amorphous thin films. *Physical Review B* **99**, 054515 (2019).

3. Saito, Y., Nojima, T. & Iwasa, Y. Quantum phase transitions in highly crystalline two-dimensional superconductors. *Nature Communications* **9**, 778 (2018).

4. Ullah, S. & Dorsey, A. T. Critical fluctuations in high-temperature superconductors and the Ettingshausen effect. *Physical Review Letters* **65**, 2066-2069 (1990).

5. Ullah, S. & Dorsey, A. T. Effect of fluctuations on the transport properties of type-II superconductors in a magnetic field. *Physical Review B* **44**, 262-273 (1991).

6. Theunissen, M. H. & Kes, P. H. Resistive transitions of thin film superconductors in a magnetic field. *Physical Review B* **55**, 15183-15190 (1997).

7. Baturina, T. I., Mironov, A. Y., Vinokur, V. M., Baklanov, M. R. & Strunk, C. Hyperactivated resistance in TiN films on the insulating side of the disorder-driven superconductor-insulator transition. *JETP Letters* **88**, 752-757 (2008).

8. Baturina, T. I., Strunk, C., Baklanov, M. R. & Satta, A. Quantum Metallicity on the High-Field Side of the Superconductor-Insulator Transition. *Physical Review Letters* **98** (2007).

9. Baturina, T. I., Mironov, A. Y., Vinokur, V. M., Baklanov, M. R. & Strunk, C. Localized Superconductivity in the Quantum-Critical Region of the Disorder-Driven Superconductor-Insulator Transition in TiN Thin Films. *Physical Review Letters* **99**, 257003 (2007).

10. Sambandamurthy, G., Engel, L. W., Johansson, A. & Shahar, D. Superconductivity-Related Insulating Behavior. *Physical Review Letters* **92**, 107005 (2004).

11. Fisher, M. P. A. Quantum phase transitions in disordered two-dimensional superconductors. *Physical Review Letters* **65**, 923-926 (1990).

12   Haviland, D. B., Liu, Y. & Goldman, A. M. Onset of superconductivity in the two-dimensional limit. *Physical Review Letters* **62**, 2180-2183 (1989).

13   Liu, Y., Haviland, D. B., Nease, B. & Goldman, A. M. Insulator-to-superconductor transition in ultrathin films. *Physical Review B* **47**, 5931-5946 (1993).

14   Yazdani, A. & Kapitulnik, A. Superconducting-Insulating Transition in Two-Dimensional $\mathit{a}$-MoGe Thin Films. *Physical Review Letters* **74**, 3037-3040 (1995).

15   Aubin, H. *et al.* Magnetic-field-induced quantum superconductor-insulator transition in $Nb_{0.15}Si_{0.85}$. *Physical Review B* **73**, 094521 (2006).

16   Yadav, S., Kaushik, V., Saravanan, M. P. & Sahoo, S. Probing electron-electron interaction along with superconducting fluctuations in disordered TiN thin films. *Physical Review B* **107**, 014511 (2023).